\documentclass[twocolumn,showpacs,preprintnumbers,prd,superscriptaddress,nofootinbib]{revtex4}
\usepackage{amsmath}
\usepackage{amsfonts}
\usepackage{amssymb}
\usepackage{graphicx}
\usepackage[titletoc]{appendix}
\usepackage{color}

\graphicspath{{Graphics/}}

\begin{document}

\title{Cosmography and constraints on the equation of state of the Universe in various parametrizations}


\author{Alejandro Aviles}
\email{aviles@ciencias.unam.mx}
\affiliation{Instituto de Ciencias Nucleares, Universidad Nacional Aut\'onoma de M\'exico, AP 70543, M\'exico, DF 04510, Mexico}
\affiliation{Departamento de F\'isica, Instituto Nacional de Investigaciones Nucleares, AP 70543, M\'exico, DF 04510, Mexico}

\author{Christine Gruber}
\email{chrisigruber@physik.fu-berlin.de}
\affiliation{Institut fuer Theoretische Physik, Freie Universitaet Berlin,
             Arnimallee 14, D-14195 Berlin, Germany}

\author{Orlando Luongo}
\email{orlando.luongo@roma1.infn.it}
\affiliation{Instituto de Ciencias Nucleares, Universidad Nacional Aut\'onoma de M\'exico, AP 70543, M\'exico, DF 04510, Mexico}
\affiliation{Dipartimento  di Fisica and Icra, Universit\`a di Roma "La Sapienza",
             Piazzale Aldo Moro 5, I-00185, Roma, Italy}
\affiliation{Dipartimento di Scienze Fisiche, Universit\`a di Napoli "Federico II",
             Via Cinthia, I-80126, Napoli, Italy}

\author{Hernando Quevedo}
\email{quevedo@nucleares.unam.mx}
\affiliation{Instituto de Ciencias Nucleares, Universidad Nacional Aut\'onoma de M\'exico, AP 70543, M\'exico, DF 04510, Mexico}
\affiliation{Dipartimento  di Fisica and Icra, Universit\`a di Roma "La Sapienza",
             Piazzale Aldo Moro 5, I-00185, Roma, Italy}

\begin{abstract}

We use cosmography to present constraints on the kinematics of the Universe,
without postulating any underlying theoretical model. To this end, we use a
Monte Carlo Markov Chain analysis to perform comparisons to the supernova Ia
Union 2 compilation, combined with the Hubble Space Telescope measurements
of the Hubble constant, and the Hubble parameter datasets.
We introduce a sixth order cosmographic parameter and show that it does not
enlarge considerably the posterior distribution when comparing to the fifth
order results. We also propose a way to construct viable parameter variables
to be used as alternatives of the redshift $z$. These can overcome both the
problems of divergence and lack of accuracy associated with the use of $z$.
Moreover, we show that it is possible to improve the numerical fits by
re-parameterizing the cosmological distances. In addition, we constrain the
equation of state of the Universe as a whole by the use of cosmography. Thus, 
we derive expressions which can be directly used to fit the equation
of state and the pressure derivatives up to fourth order. To this end, it is 
necessary to depart from a pure cosmographic analysis and to assume the Friedmann 
equations as valid. All our results are consistent with the $\Lambda$CDM model, 
although alternative fluid models, withnearly constant pressure and no 
cosmological constant, match the results accurately as well.
\end{abstract}

\pacs{98.80.-k, 98.80.Jk, 98.80.Es}
\maketitle

\section{Introduction}

Ever since the pioneering works of two separate groups in
1998~\cite{SNeIa-1}, cosmological observations indicate a late time
accelerated Universe.  More recently, additional evidence coming from
other experiments~\cite{SNeIa-2,SNeIa-3} confirms that the
Universe is going through an accelerated expansion. Thus, the
existence of the acceleration is assumed to be a consolidated
feature of cosmology~\cite{SNeIa-4}. Unfortunately, the physical
mechanism from which this cosmic speed up originates is still
unclear; the common way to deal with this is to assume, in addition
to the standard matter term, the existence of a further exotic fluid
which influences
the dynamics of the Universe~\cite{orly1,orly1a,orly1b,orly1c}. Due
to the lack of knowledge on the physical nature of this fluid, we
usually refer to it as dark energy (DE). So far, DE is only
observationally witnessed, while the micro-physics behind
it remains totally
undisclosed~\cite{orly2,orly2a,orly2b,orly2c,orly2d}.  One of the
most dubious properties of DE is that it exhibits a negative equation of state (EoS) parameter,
counteracting the attractive action of gravity~\cite{EoS}. The need
of a negative pressure hints at the non-baryonic nature of DE, since no
common matter is expected to show such a property. Besides that, the
total amount of cosmological matter in the Universe appears to be
dominated by a non-baryonic (cold) dark matter (DM) component,
which accounts for about $23\%$ of the total energy content of the
Universe.  On the other hand, the common
baryonic matter in the Universe only accounts for $4\%$ of the whole
energy content. This shows that the standard visible matter is
actually not enough to guarantee the stability of structure observed
at different astrophysical and cosmological scales, and DM cannot be ignored for the
dynamics of the whole Universe~\cite{DM,alejandrowork}.

Consequently, our knowledge of the correct cosmological model seems
to be lacking of some ingredients. However, in order to investigate the
effects of DE and DM in Einstein's equations, one introduces a common
energy momentum tensor with a pressureless term, i.e. $P_m=0$,
describing the total visible and non-visible matter content, and an
additional term with a negative pressure to represent DE
\cite{libro}. Together with these assumptions, one generally
considers a homogeneous and isotropic Universe, depicted by the
Friedmann-Robertson-Walker (FRW) metric,
$ds^2=-c^2dt^2+a(t)^2(dr^2/(1-kr^2)+r^2\sin^{2}\theta
d\phi^2+r^2d\theta^2)$. In addition, observations of the large scale geometry
of the Universe suggest a spatially quite flat Universe, so hereafter we will assume $k=0$.
To account for the effects of DE, the
simplest and most tested assumption deals with the introduction of a
cosmological constant term $\Lambda$ into the Einstein equations.
According to quantum field theory, the constant is interpreted as a
vacuum energy contribution, and naturally leads to a negative EoS parameter
with a positive energy density and negative pressure. The
corresponding model, which is straightforwardly derived by solving
the Einstein equations with the cosmological constant, is known as
$\Lambda$CDM~\cite{paddy}, a model which by now achieved the
status of the standard cosmological model. The reason which
induces cosmologists to assume this particular model to be the
standard one is that it excellently fits all observational data
with high precision~\cite{Weinberg2008,WMAP7}. Moreover, it is only
relying on a remarkably small number of cosmological parameters,
without any \emph{ad hoc} additional terms~\cite{Tsuji2010}.
Unexpectedly, observations show that both the magnitudes of
matter and $\Lambda$ are
comparable at our time. Indicating with $\Omega_{\text{m}}$ and
$\Omega_\Lambda$ the magnitudes of matter and DE respectively,
observational bounds show that
$\Omega_{\Lambda}/\Omega_{\text{m}}\approx 2.7$. This feature implies
a strange and unexpected coincidence problem --because DE is
expected to evolve separately from matter, it is quite astonishing
to imagine that near the present time the two magnitudes
should be so close to each other.

On the other hand, another uncomfortable shortcoming plagues the
standard model. The observational limits on the magnitude of the
cosmological constant disagree with the predicted value for about
$10^{123}$ orders of magnitude, leading to a serious fine-tuning
problem~\cite{Weinberg1989}. This deeply disturbs the otherwise
appealing picture of a cosmological constant and, together with the
coincidence problem, dramatically afflicts the standard cosmological
paradigm. Despite its success in explaining the observational
data, the $\Lambda$CDM model is therefore theoretically
incomplete\footnote{Moreover, nearly all the extensions of it appear
to fail as well. For a recent and mentionable alternative, which
naturally extends $\Lambda$CDM in general relativity, conforming to
all the experimental bounds, see~\cite{alejandrowork, Balbi07, Luongo2011}.}, or at least
not well understood. Motivated by these defects, a \emph{mare
magnum} of different models has been proposed during the last
decades; as a short sample see~\cite{orly3_1} and references therein.
In this work we wonder whether the cosmological constant must be considered as
the real unique explanation of DE, or if there exists a hidden
mechanism behind the nature of the cosmic speed up. An enticing way
to understand if $\Lambda$CDM is the favorite candidate for DE is
represented by the analyses through model independent tests. Such
procedures should be able to disclose the fundamental nature of DE
without postulating a certain model \emph{a priori}. In this way, it
would be possible to analyze the dynamics of the Universe without
imposing a cosmological constant from the beginning. If $\Lambda$
really exists, no significant deviations from a constant EoS must be
found by model-independent tests. As a consequence, it is
necessary to inquire how much of modern cosmology is really
independent of the Friedmann
equations~\cite{CattViss2005,visserrimo}. In other words,
distinguishing between kinematics and dynamics is viewed as a tool
to discriminate fairly among models, in order to reveal the correct
cosmological paradigm.

Surely one of the most powerful model independent approaches is
represented by cosmography~\cite{CattViss2008}. Cosmography,
sometimes also referred to as cosmo-kinetics, was first discussed by
Weinberg in~\cite{Weinberg2008} and then extended by Visser
in~\cite{CattViss2005}. The underlying philosophy of cosmography is
to involve the cosmological principle only. So, the FRW metric is
the only ingredient that cosmography uses for obtaining bounds on
the observable Universe. Cosmography permits us to infer how much DE
or alternative components are required in regard to satisfy the
Einstein equations. The idea is to expand some observables such as
the cosmological distances or the Hubble parameter, into power
series, and relating cosmological parameters directly to these
observable quantities. In doing so, it is possible to appraise which
models behave fairly well and which ones should be discarded as a
consequence of not satisfying the basic demands introduced by
cosmography. So cosmography strives for the development of a
procedure able to constrain the kinematics of the Universe.

In this paper, we present an extension of the promising approach
debated in Visser et al.~\cite{CattViss2008}; we devote our efforts
both to constraining $\Lambda$CDM and investigating whether it is
the only possibility to explain the cosmological acceleration, or
whether there are prominent alternatives~\cite{nobel}. In
particular, we adopt the idea of cosmography developed
in~\cite{CattViss2005,CattViss2008,CattViss2007,Luongo,Xu,capozzie}
and we improve it, by assuming an extended class of fitting
quantities, assembled by a number of different cosmological
distances. For theoretical reasons, which we will discuss in the
next sections, we introduce new parameterizations of the redshift
variable besides $z$, in order to improve the fitting procedure.
These parameterizations are designed to reduce the problems
associated to the experimental analysis at redshift $z>1$. Thence,
we make use of the most recent data of the Union 2 supernovae Ia
(SNeIa), of the Hubble Space Telescope (HST) measurements of the
Hubble factor, and of the
$H(z)$ compilations~\cite{tutti e tre}, through a Markov Chain Monte
Carlo (MCMC) method, by modifying the publicly available code CosmoMC
\cite{codice}. Afterwards, we also include a parametrization of the
cosmological distances in terms of the EoS of the Universe as a whole and of
their pressure derivatives. This allows us to directly fit the EoS of
the Universe without having to undergo disadvantageous error
propagation. In doing this, we assume the validity of the cosmological principle, and of General
Relativity, since for this analysis it is necessary to invoke the Friedmann
equations. This gives us certain constraints on the EoS and on the pressure
derivatives in the framework of General Relativity.

The paper is organized as follows. In Sec.~II we discuss the physics
behind the concept of cosmography and its implications for modern
cosmology in more detail; we introduce the new cosmographic
coefficient $m$, and discuss how to build up a viable alternative
parametrization to the redshift $z$. We then propose three new
parameterizations and we study their properties, in view of the
fitting procedure. In Sec.~III we apply these recipes and present
the results of the cosmographic fits by a numerical MCMC analysis.
Section~IV deals with the concept of the EoS; in particular we
relate the EoS and the derivatives of pressure to the luminosity
distance, in order to carry out a direct fit to cosmological data.
We use these results to derive constraints on $\Lambda$CDM and the
most generic DE model, characterized by the function $G(z)$,
reducing to $G(z=0)=1-\Omega_m$ at $z=0$. In Sec.~V finally we draw
conclusions and give an outlook to further paths of investigations.

\section{The role of Cosmography}

In this section we focus on the role of cosmography in modern
cosmology. Its aim is the study of the kinematic quantities,
characterizing the cosmological scenario. For this reason,
cosmography is also called \emph{cosmo-kinetics}, or kinematics of
the Universe. We therefore limit ourselves to the
smallest number of assumptions possible. First, we presume the
validity of the cosmological principle. Second, we suppose that the
EoS of the Universe is determined by a non-specified number of
different cosmological fluids~\footnote{These fluids include matter, radiation, 
curvature, dark energy and so forth.}. We assume that the total pressure of these fluids, namely $P$,
can be written as $P=\sum_iP_i$, and its total EoS parameter
$\omega = \sum_iP_i/\sum_i\rho_i $. Here the index $i$ runs over all the
involved cosmological fluids. Following
these assumptions, the paradigm of cosmography was first developed
by Weinberg~\cite{Weinberg2008}, who proposed to expand the scale
factor in terms of a Taylor series around the present time
$t_0$~\footnote{Instead of the scale factor, also the Hubble parameter 
or the luminosity distance could be expanded.
At late times it is allowed to neglect radiation within the energy
momentum tensor in Einstein's equations. In addition, as will be
clarified later, we superimpose a spatially flat geometry, in
accordance with the most recent observations.}. Following these
recipes, it is natural to expect that many other physical quantities
of interest, apart from $a(t)$, can be expanded as well. The power
series coefficients in the expansion of the scale factor are known
in the literature as cosmographic series (CS), when evaluated at our
time $t_0$; these quantities are related to the scale factor
derivatives.

A feasible consequence of this prescription is that cosmography does
not depend on the choice of a cosmological model. As a matter of
fact, almost all cosmological tests assume \emph{a priori} that
the model under consideration is statistically favored; however,
this creates a degeneracy among models and often it remains
difficult to understand which model is really favored. Cosmography
is, among various cosmological tests, one of the ways to alleviate
that degeneracy. However, although cosmography is reviewed as a
model independent procedure, a few words should be spent regarding
the role of the spatial curvature, $k$. In particular, modern
cosmological data are not enough at present to fix stringent
convergence limits on the CS and $k$. It is possible to show that
the term proportional to the (present) ``variation of acceleration",
i.e. $j_0$, cannot be measured alone. Defining the curvature density
as $\Omega_k\equiv\Omega_0-1$, where $\Omega_0$ represents the total
density of the Universe, then one measures $j_0+\Omega_0$
\cite{Faber2006}. Motivated by WMAP 7 results~
\cite{WMAP7}, we propose here to restrict the analysis to the
spatially flat case, in which $k=0$. This
naturally overcomes the dependence of $j_0$ on $\Omega_0$, letting
cosmography be independent of any particular cosmological
framework.

Now we have all the ingredients to expand the scale factor into a series,
yielding

\begin{eqnarray}
a(t) & = & a_0\cdot \Bigg[ 1 + \frac{da}{dt}\Big|_{t_0} (t-t_0) \nonumber\\
&+&\frac{1\,}{2!}\frac{d^2a}{dt^2}\Big|_{t_0} (t-t_0)^2 + \frac{1\,}{3!} \frac{d^3a}{dt^3}\Big|_{t_0} (t-t_0)^3 \nonumber\\
&+&\frac{1\,}{4!} \frac{d^4a}{dt^4}\Big|_{t_0} (t-t_0)^4 + \frac{1\,}{5!} \frac{d^5a}{dt^5}\Big|_{t_0} (t-t_0)^5 \nonumber\\
&+&\frac{1\,}{6!} \frac{d^6a}{dt^6}\Big|_{t_0} (t-t_0)^6 +\mathcal{O}((t-t_0)^7) \Bigg]\,,
\label{eq:expa}
\end{eqnarray}

where we truncated the series at the sixth order in $\Delta t\equiv
t-t_0$. Here, we assume that $t-t_0>0$.
Moreover, the constant $a_0$ is the scale factor evaluated today. Without loss of generality, it is licit to identify
hereafter $a_0=1$. Equation~($\ref{eq:expa}$) can be recast as

\begin{eqnarray}\label{serie1a}
a(t) & = &  1 - H_0 \Delta t - \frac{q_0}{2} H_0^2 \Delta t^2 -
\frac{j_0}{6} H_0^3 \Delta t^3 +
   \frac{s_0}{24} H_0^4 \Delta t^4 \nonumber\\
   &&-\, \frac{l_0}{120} H_0^5 \Delta t^5 + \frac{m_0}{720} H_0^6 \Delta t^6 +
   \mathcal{O}(\Delta t^7)\,,
\end{eqnarray}

with the definition of the cosmographic coefficients as

\begin{eqnarray}\label{pinza}
H \equiv \frac{1}{a} \frac{da}{dt}\,,\quad&\,& \quad q \equiv -\frac{1}{aH^2} \frac{d^2a}{dt^2}\,,\nonumber\\
j  \equiv \frac{1}{aH^3} \frac{d^3a}{dt^3}\,,\quad&\,& \quad s \equiv \frac{1}{aH^4} \frac{d^4a}{dt^4}\,,\\
l  \equiv \frac{1}{aH^5} \frac{d^5a}{dt^5}\,,\quad&\,& \quad m \equiv
\frac{1}{aH^6}\frac{d^6a}{dt^6}\,. \nonumber
\end{eqnarray}

Having an expansion for $a(t)$ is equivalent to having an expansion
of the redshift $z$, in terms of $H_0 \Delta t$ (see~\cite{cinesi
solo senza mio}).

As previously stressed, Eqs.~($\ref{pinza}$), if evaluated at our
time, are referred to as the CS. The subscript $``0"$ in Eq.
($\ref{serie1a}$) indicates that the coefficients are evaluated at
$t=t_0$. In particular, each term has its own specific physical
interpretation. For example, $q$, the so-called acceleration
parameter, specifies whether the Universe is accelerating or
decelerating, depending on the sign. An accelerating Universe leads to
$-1\leq q_0<0$. On the contrary, a positive $j_0$ implies that $q$
changes sign as the Universe expands, and so forth for all the rest
of the parameters. We usually attribute the names of jerk and snap to
$j$ and $s$ respectively; so far, no universal name is associated to
$l$. Here, we additionally introduce $m$ as a further higher order
term. It is useful to combine the CS among themselves and express
them in terms of each other, yielding

\begin{eqnarray}\label{eq:CSoftime}
q&=&-\frac{\dot{H}}{H^2} -1\,, \nonumber\\
j&=&\frac{\ddot{H}}{H^3}-3q-2\,, \nonumber\\
s&=&\frac{H^{(3)}}{H^4}+4j+3q\left(q+4\right)+6\,,\\
l&=&\frac{H^{(4)}}{H^5}-24 - 60 q - 30 q^2 - 10 j \left(q+2\right) + 5 s\,, \nonumber\\
m&=&\frac{H^{(5)}}{H^6} + 10 j^2 + 120 j \left(q+1\right)
+\nonumber\\ &&3 \left[2 l + 5 \left(24 q
 + 18 q^2 + 2 q^3 - 2 s - q s+8\right)\right]\,. \nonumber
\end{eqnarray}

Here, the dots and the numbers in brackets indicate the derivatives
with respect to the cosmic time. By converting the derivatives
in Eq.~\eqref{eq:CSoftime} from time to redshift, and inverting
the relations, we can obtain the Hubble parameter as an expansion
in terms of redshift, $H(z)$ (the results can be found in
Appendix~\ref{app:Hubble}).
Analogously, we can also expand
other observable physical quantities in order to fit the cosmological
data with the obtained functions.

It would be interesting to expand
commonly used notions of cosmological distances to the same order of
the Taylor expansion as the scale factor before.
To this end, let us now introduce several examples of
distances between two objects in cosmology, following the
prescriptions given in~\cite{CattViss2007}. Here, we
state the luminosity distance $d_L$ and other four
alternative distances, namely the photon flux distance $d_F$, the
photon count distance $d_P$, the deceleration distance $d_Q$ and the
angular diameter distance $d_A$. These distances are defined as

\begin{eqnarray}\label{serie2}
    d_L & = & a_0 r_0 (1+z) = r_0 \cdot \frac{1}{a(t)}\,, \nonumber \\
    d_F & = & \frac{d_L}{(1+z)^{1/2}} = r_0 \cdot \frac{1}{\sqrt{a(t)}}\,, \nonumber \\
    d_P & = & \frac{d_L}{(1+z)} = r_0\,, \\
    d_Q & = & \frac{d_L}{(1+z)^{3/2}} = r_0 \cdot \sqrt{a(t)}\,, \nonumber \\
    d_A & = & \frac{d_L}{(1+z)^2} = r_0\cdot a(t)\,.\nonumber
    \label{eq:dist}
\end{eqnarray}

The last four notions of distances are less commonly used in
literature.  Besides the luminosity distance $d_L$, which gives the
ratio of the apparent and the absolute luminosity of an
astrophysical object, we consider the photon flux distance
$d_F$, which is not calculated from the energy flux in the detector,
but from the photon flux, which is experimentally easier to measure.
The photon count distance $d_P$ is based on the total number of
photons arriving at the detector as opposed to the photon rate. The
so-called deceleration distance $d_Q$ has been introduced
in~\cite{CattViss2005} without having an immediate physical meaning,
but in return a very simple and practical dependence on the
deceleration parameter $q_0$. Finally, the angular diameter distance
$d_A$ was defined in~\cite{Weinberg2008} as the ratio of the
physical size of the object at the time of light emission and its
angular diameter observed today. To completely determine the
distance expansions, we still need to calculate $r_0$. It is defined
as the distance $r$ a photon travels from a light source at $r=r_0$
to our position at $r=0$. It is defined as\footnote{Here we have
omitted a factor of $c$ in the numerator; for now and for the rest
of the theoretical calculations in this paper, we will assume
$c=1$.}

\begin{equation}
r_0 = \int_{t}^{t_0}{\frac{dt'}{a(t')}}\,.
\end{equation}

We can calculate this quantity by inserting the power series
expansion for the inverse of the scale factor and integrating each
term in the sum separately. Finally the results are used to complete
the calculations of the cosmological distances in terms of the
redshift $z$. We report in Appendix~\ref{app:z} the expansions of
all the distances in Eqs.~($\ref{serie2}$) in terms of $z$.
These results can be
compared with those of~\cite{CattViss2008}, in which the authors
truncated the series at a lower order. In particular, they claimed
the need of using all the distances for a cosmographic test. In
principle, this may be true, because all the various cosmological
distances rely on the fundamental assumption that the total number
of photons is conserved on cosmic scales. Hence, there is no reason
to discard one distance for another one, since all of them
fulfill this condition. Unfortunately, there exists a \emph{duality
problem} plaguing such distances
\cite{hungoranzaolonyuinokretreajfbsoauyebvvvqoaufhs}. This problem
is so far an open question of observational cosmology~\cite{bll}. On
the other hand, it has been suggested that the luminosity distance
$d_L$ is fairly well adapted to the cosmological data used in
combined tests with supernovae Ia and
HST~\cite{IWILLCITE,union2,bll2}. Even though this topic is still
object of debate~\cite{bll3}, unlike Cattoen and
Visser~\cite{CattViss2008} we limit our attention to $d_L$ only. We
motivate this choice with the above considerations on the good
adaptation of $d_L$ to the data, and with its general use in
literature~\cite{bll4}.

There are two main problems arising in the context of cosmography.
In principle, the Taylor series is expected
to diverge at $z\geq 1$. This is a consequence of the fact that we
are expanding around $z\sim0$ and so when $z>1$, we get problems with
convergence. Moreover, the finite truncations we made
represent only an approximation of the exact function, giving
therefore possibly misleading results. Thus, while the second
problem can be alleviated by expanding to
higher orders in adding more coefficients, this measure can
introduce divergences into the analysis.
These issues are intimately connected to the
problem of systematic errors. In fact, if errors are large enough, it is
possible that bad convergence may afflict the numerical results.

We improve the accuracy of our work by using the Union 2
compilation, which reduces the problem of systematics,
easing the second problem. On the other hand, in order to
overcome the first issue, different parameterizations of the
fitting functions can be taken into account.
The idea is to carry out the expansion with a different variable
which is used \emph{ad interim}
and is constructed to be limited in a more stringent interval. While
$z\in[0,\infty]$, a new variable should for example be restricted
to the interval $[0,1]$.

A fairly well-known possibility is represented by the variable
\begin{equation}\label{hj}
y_1 = \frac{z}{1+z}\,,
\end{equation}
frequently used in literature~\cite{CattViss2005,Cap_Lazkoz}. The limits in the
past Universe, i.e. $z \in [0,\infty]$, read $y_1\in [0,1]$, while
in the future, i.e. $z \in [-1,0]$:  $y_1\in [-\infty,0]$.
Immediately we notice that $y_1$ can be expanded as $z$ before as
$y_1=y_1(H_0\Delta t)$; then, it is feasible to invert it, having
$H_0 \Delta t$ in terms of $y_1$. Then, we can express the distances
as functions of $y_1$ (for the results, see Appendix~\ref{app:y1}).

Furthermore, we propose a way to construct other viable parameterizations
of the redshift variable. To this end, we introduce below three new
propositions, namely $y_2,y_3$ and $y_4$, as

\begin{eqnarray}\label{ilbosco}
y_2&=&\arctan{\Bigl(\frac{z}{z+1}\Bigr)} = \arctan\left(1-a\right)\,,\nonumber\\
y_3&=&\frac{z}{1+z^2}\,,\\
y_4&=&\arctan{z}\,,\nonumber
\end{eqnarray}

whose limits are, for $z \in [0,\infty]: y_2\in [0,\frac{\pi}{4}],
y_3\in [0,0], y_4\in [0,\frac{\pi}{2}]$ and $z \in [-1,0]: y_2\in
[\frac{\pi}{2},0], y_3\in [-\frac{1}{2},0], y_4\in
[-\frac{\pi}{4},0]$ and in which we used the definition of the
scale factor, i.e. $a\equiv(1+z)^{-1}$.

We adopted the $\arctan$ in the parameterizations of $y_{2,4}$ because it
behaves smoothly and it is suited to give
well-defined limits at $z \rightarrow \infty$. On the contrary,
$y_3$ is a polynomial in $z$; so apparently, we would not expect it
to lead to significantly different fitting behavior, but just to represent an
alternative worth investigating. \\
Equations~($\ref{ilbosco}$) can be expanded into Taylor series for
$z\ll1$, and then be inverted for $y_{2,3,4}$ in order to give an
expression for $H_0 \Delta t (y_{2,3,4})$, up to sixth order. Then
the distances as functions of $y_{1,2,3,4}$ can be calculated as
well (results see Appendices~\ref{app:y2}-\ref{app:y4}).

\begin{figure}
   \begin{center}
   \includegraphics[width=0.5\textwidth]{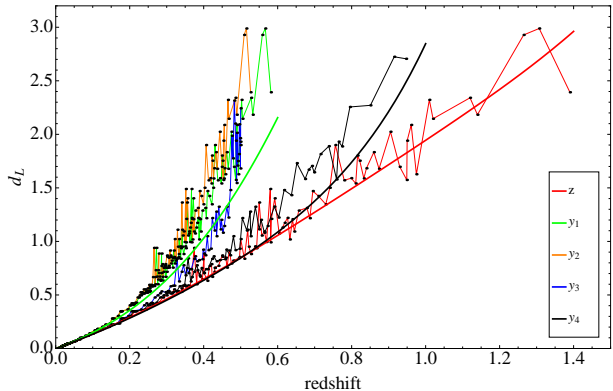}
  {\small \caption{(color online) Luminosity distance (in units of
    $10^{26}\,$m), over different redshifts $z$ (red), $y_1$ (green),
    $y_2$ (orange), $y_3$ (blue) and $y_4$ (black).}
  \label{fig:redshifts}}
  \end{center}
\end{figure}

By definition, all these parameterizations are built up to avoid
divergences at $z>1$. Thus, one can wonder whether all of them turn
out to be equally suitable for constraining the CS. The answer can
be partly predicted by comparing the supernova data of the
luminosity distance, as in Fig.~\ref{fig:redshifts}, for $z$ and
$y_{1,2,3,4}$. The worst example is clearly the redshift $y_3$. Its
definition suggests that it scales down more quickly, compared to
the redshift $z$, as can also be seen in Fig.~\ref{fig:redshifts}. This
means that a region of $z\in [0,1.5]$ is reduced to a much
smaller interval $y_3\in [0,0.5]$. Thereafter, we expect that, when
the curve bends too quickly, the fits become more difficult. It
follows that, as the curve trends become more extreme, a suppression
of lower redshifts to the advantage of higher ones can occur. In
other words, $z\in[0,0.5]$ weighs less than $z\geq 0.5$; therefore,
we guess that $y_3$ would work better if all the cosmological data
were for $z\gg 1$.

As it can be seen in Fig.~\ref{fig:redshifts}, the luminosity
distance curves of data points over redshift are slightly flexed,
becoming steeper towards higher redshifts. Also the redshifts $y_1$,
$y_2$ and $y_4$ lead to steeper curves than $z$, however, redshift
$y_3$ behaves the most extreme. According to the cited criteria,
$y_3$ is the least suitable of redshift notions. This conjecture is
also backed by the fitting results, which confirm that $y_3$
does not work well in the application to SNeIa data. Another
disadvantage of $y_3$ is that it does not have a uniquely defined
inverse. For these reasons, we decided to take it out of the
analysis. Through similar arguments, we remove the second-worst
redshift, $y_2$, as well. Summing up, in order to contrive a viable
redshift parametrization, the following conditions must be
satisfied:
\begin{enumerate}
 \item The luminosity distance curve should not behave too
steeply in the interval $z<1$.
 \item The luminosity distance curve should not exhibit sudden flexes.
 \item The curve should be one-to-one invertible.
\end{enumerate}
From Fig.~\ref{fig:redshifts}, we notice that
the last introduced redshift $y_4$, although still producing a
steeper curve than $z$, is expected to work better than $y_1$. Thus,
the rest of the analysis, including further calculations and
fittings, is carried out for the redshifts $z$, $y_1$ and $y_4$.

\section{The fitting procedure and the cosmographic results}

In Sec.~II, we explained how to construct viable cosmological
parameterizations to alleviate the problems associated to cosmography. To
this end, we introduced $y_4$ and we investigated its theoretical
viability to fit the cosmological data. We now have all the
ingredients to develop a MCMC procedure to find
cosmographic constraints for the CS, using the three redshifts $z$,
$y_1$ and $y_4$ for the fitting analysis. We will include the
sixth order of the CS, i.e. $m_0$, and particularly focus on the
following aspects:

\begin{itemize}
\item We investigate whether redshift $y_4$ is actually suitable to obtain
accurate values of the CS, as theoretically predicted, and
we explore the ranges of low and higher redshifts;

\item We analyze how well the introduced CS parameter $m_0$ can be
constrained, in particular, we examine whether its introduction
significantly enlarges the dispersion of the estimation of the other
parameters;

\item We find out if the concordance model, i.e. $\Lambda$CDM, is in agreement
with the cosmographically found numerical results.
\end{itemize}

For our purposes, we use the data of the SNeIa Union 2 compilation
by the supernovae cosmology project~\cite{union2}. We also adopt the
HST measurements on 600 Cepheides, which
impose a Gaussian prior on the Hubble parameter today of $H_0 = 74.0
\pm 3.6 \,\text{km/s/Mpc}$~\cite{Riess2009}, and the measurements of
the Hubble parameter $H(z)$ at twelve different redshifts ranging
from $z=0.1$ to $z=1.75$~\cite{Stern:2009ep}. We divide our analysis
into two sets of observations, namely set 1, which comprises
Union 2 together with HST, and set 2, being the Union 2 dataset
with both HST and $H(z)$ measurements. For the sake of completeness,
it is in order to cite recent works using gamma ray bursts (GRBs)
as possible distance indicators~\cite{Vitagliano:2009et,Luongo,Xu}.
However, considering GRBs in such analyses is mere speculation,
since GRBs are not standard candles~\cite{Schaefer:2006pa}. This
approach seems to result in wrong estimations, or at least
inadequate results. It will be shown that our results differ from
those obtained by using GRBs; we will show a set of results in
better agreement with $\Lambda$CDM than those obtained by using
GRBs, which points to an inadequacy of the use of GRBs in
cosmography. We also  exclude observation
data from baryonic acoustic oscillations from our analysis;
we deem that introducing baryon acoustic oscillations means reducing the
model-independence of the whole analysis~\cite{Durrer2011}.

In the following, we will use the CS combined together
in three sets with different maximum order of
parameters:

\begin{eqnarray}
 \mathcal{A} &=& \{ H_0, q_0, j_0, s_0 \}\,, \nonumber\\
 \mathcal{B} &=& \{ H_0, q_0, j_0, s_0, l_0 \}\,, \\
 \mathcal{C} &=& \{ H_0, q_0, j_0, s_0, l_0, m_0 \}\,.\nonumber
\end{eqnarray}

We expect a slower convergence of the last dataset, since the
introduction of $m_0$ can decrease the accuracy of convergence. To
constrain the parameters, we use a Bayesian technique in which the
best fits are those maximizing the likelihood function $\mathcal{L}
\propto \exp (-\chi^2/2 )$. Since the different observations are not
correlated, the function $\chi^2$ is simply given by the sum $\chi^2
= \chi_{\text{Union 2}}^2 + \chi_{\text{HST}}^2 +\chi_{H(z)}^2$. We
explore the space of parameters with a MCMC approach, modifying the
publicly available code CosmoMC~\cite{codice}. We do the analysis
for the three sets' parameter space, for the two sets of
observations, and for each of the three considered redshifts.
Accordingly, we perform 18 different constraint parameter analyses.
To obtain the posterior samples we assume flat priors over the
intervals $-6<q_0<6$, $-20<j_0<20$, $-200< s_0< 200$,
$-500<l_0<500$, and $-3000<m_0<3000$.

In Tables~\ref{table:z},~\ref{table:y1}, and~\ref{table:y4} we show
the best fits and their $1\sigma$-likelihoods for the  redshifts
$z$, $y_1$ and $y_4$ respectively.

\begin{table*}
\caption{{\small Table of best fits and their likelihoods (1$\sigma$) for redshift $z$, for the three sets of
         parameters $\mathcal{A}$, $\mathcal{B}$ and $\mathcal{C}$.
                 Set 1 of observations is Union 2 + HST. Set 2 of observations is Union 2 + HST + $H(z)$.}}

\begin{tabular*}{\textwidth}{@{\extracolsep{\fill}}ccccccc}
\hline\hline

{\small Parameter}  &   {\small $\mathcal{A}$, Set 1}                  & {\small $\mathcal{A}$, Set 2 }
                    &   {\small $\mathcal{B}$, Set 1}                  & {\small $\mathcal{B}$, Set 2 }
                    &   {\small $\mathcal{C}$, Set 1}                  & {\small $\mathcal{C}$, Set 2 }  \\ [1.5ex]
                    &   {\small $\chi^2_{min} = 530.1\,{}^b$ }              & {\small 545.6 }
                    &   {\small 530.1 }                                     & {\small 544.5 }
                    &   {\small 530.0 }                                     & {\small 544.3 } \\[0.8ex]  
\hline \
{\small$H_0$}       & {\small $74.35$}{\tiny${}_{-7.50}^{+7.39}$}           & {\small $74.22$}{\tiny ${}_{-5.08}^{+5.23}$}
                    & {\small $73.77$}{\tiny ${}_{-7.35}^{+8.36}$}          & {\small $74.20$}{\tiny ${}_{-5.49}^{+5.01}$}
                    & {\small $73.72$}{\tiny ${}_{-7.12}^{+8.47}$}          & {\small $73.65$}{\tiny ${}_{-5.35}^{+5.92}$} \\[0.8ex]

{\small $q_0$}      & {\small $-0.7085$}{\tiny ${}_{-0.5952}^{+0.6074}$}    & {\small $-0.6149$}{\tiny ${}_{-0.2238}^{+0.2716}$}
                    & {\small $-0.6250$}{\tiny ${}_{-0.4953}^{+0.5580}$}    & {\small $-0.6361$}{\tiny ${}_{-0.3645}^{+0.3720}$}
                    & {\small $-0.6208$}{\tiny ${}_{-0.6773}^{+0.4849}$}    & {\small $-0.5856$}{\tiny ${}_{-0.3445}^{+0.3884}$} \\[0.8ex]

{\small $j_0$}      & {\small $1.605$}{\tiny${}_{-4.481}^{+6.738}$}         & {\small $1.030$}{\tiny ${}_{-1.001}^{+0.722}$}
                    & {\small $0.392$}{\tiny ${}_{-4.511}^{+4.585}$}        & {\small $0.994$}{\tiny ${}_{-2.665}^{+1.904}$}
                    & {\small $-1.083$}{\tiny ${}_{-2.218}^{+8.359}$}       & {\small $-0.117$}{\tiny ${}_{-1.257}^{+3.621}$} \\[0.8ex]

{\small $s_0$}      & {\small $2.53$}{\tiny${}_{-10.45}^{+60.61}$}          & {\small $0.16$}{\tiny ${}_{-1.03}^{+1.45}$}
                    & {\small $-5.59$}{\tiny ${}_{-34.55}^{+33.74}$}        & {\small $-1.47$}{\tiny ${}_{-10.72}^{+4.20}$}
                    & {\small $-25.52$}{\tiny ${}_{-10.90}^{+65.60}$}       & {\small $-7.71$}{\tiny ${}_{-7.83}^{+14.77}$} \\[0.8ex]

{\small $l_0$}      & {\Large --}                                           & {\Large --} 
                    & {\small $-3.50$}{\tiny ${}_{-89.19}^{+196.09}$}       & {\small $4.47$}{\tiny ${}_{-8.47}^{+41.53}$}
                    & {\small $N.C.$}                                       & {\small $8.55$}{\tiny ${}_{-27.86}^{+23.39}$} \\[0.8ex]

{\small $m_0$}      & {\Large --}                                           & {\Large --} 
                    & {\Large --}                                           & {\Large --} 
                    & {\small $N.C.$}                                       & {\small $71.93$}{\tiny ${}_{-315.76}^{+382.17}$} \\[0.8ex]

\hline \hline

\end{tabular*}

{\tiny Notes.

a. $H_0$ is given in Km/s/Mpc.

b. $N.C.$ means the results are not conclusive. The data do not constrain the parameters sufficiently.

}

\label{table:z}
\end{table*}


\begin{table*}
\caption{{\small Table of best fits and their likelihoods (1$\sigma$) for redshift $y_1$, for the three sets of
         parameters $\mathcal{A}$, $\mathcal{B}$ and $\mathcal{C}$.
                 Set 1 of observations is Union 2 + HST. Set 2 of observations is Union 2 + HST + $H(z)$.}}

\begin{tabular*}{\textwidth}{@{\extracolsep{\fill}}ccccccc}
\hline\hline
{\small Parameter}  &   {\small $\mathcal{A}$, Set 1}                  & {\small $\mathcal{A}$, Set 2 }
                    &   {\small $\mathcal{B}$, Set 1}                  & {\small $\mathcal{B}$, Set 2 }
                    &   {\small $\mathcal{C}$, Set 1}                  & {\small $\mathcal{C}$, Set 2 }  \\ [1.5ex]
                    
                    &   {\small $\chi^2_{min} = 530.1$}                   & {\small 550.1 }
                    &   {\small 529.9 }                                   & {\small 544.5 }
                    &   {\small 530.0 }                                   & {\small 545.1 } \\[0.8ex]
\hline {\small $H_0$}      & {\small $74.05$}{\tiny${}_{-7.19}^{+7.90}$}  & {\small $75.25$}{\tiny${}_{-4.87}^{+4.72}$}
                    & {\small $73.68$}{\tiny ${}_{-6.94}^{+7.77}$}        & {\small $73.30$}{\tiny ${}_{-5.22}^{+5.59}$}
                    & {\small $73.91$}{\tiny ${}_{-6.97}^{+7.60}$}        & {\small $74.49$}{\tiny ${}_{-5.59}^{+5.07}$} \\[0.8ex]

{\small $q_0$}      & {\small $-0.6633$}{\tiny${}_{-0.6580}^{+0.5753}$}   & {\small $-0.4106$}{\tiny${}_{-0.5774}^{+0.2919}$}
                    & {\small $-0.0004$}{\tiny ${}_{-1.6617}^{+0.2513}$}  & {\small $-0.2652$}{\tiny ${}_{-0.7977}^{+0.5071}$}
                    & {\small $-0.5360$}{\tiny ${}_{-0.8965}^{+0.8468}$}  & {\small $-0.4624$}{\tiny ${}_{-0.8391}^{+0.5804}$} \\[0.8ex]

{\small $j_0$}      & {\small $1.268$}{\tiny${}_{-4.273}^{+6.986}$}       & {\small $-7.746$}{\tiny ${}_{-2.252}^{+15.526}$}
                    & {\small $-13.695$}{\tiny ${}_{-1.703}^{+30.901}$}   & {\small $-7.959$}{\tiny ${}_{-5.228}^{+13.529}$}
                    & {\small $-1.646 $}{\tiny ${}_{-8.345}^{+11.637}$}   & {\small $-1.862$}{\tiny ${}_{-5.397}^{+11.021}$} \\[0.8ex]

{\small $s_0$}      & {\small $1.21$}{\tiny${}_{-9.24}^{+61.24}$} & {\small $-88.91 $}{\tiny ${}_{-11.08}^{+57.62}$}
                    & {\small $-180.95 $}{\tiny ${}_{-18.93}^{+331.75}$}  & {\small $-112.63$}{\tiny ${}_{-82.53}^{+156.60}$}
                    & {\small $-30.97$}{\tiny ${}_{-43.47}^{+90.96}$}       & {\small $-16.95 $}{\tiny ${}_{-38.79}^{+73.68}$} \\[0.8ex]

{\small $l_0$}      & {\Large --} & {\Large --} 
                    & {\small $N.C. $}                                    & {\small $N.C.  $}
                    & {\small $N.C. $}                                    & {\small $N.C.  $} \\[0.8ex]

{\small $m_0$}      & {\Large --} & {\Large --} 
                    & {\Large --}                                         & {\Large --} 
                    & {\small $N.C. $}                                    & {\small $N.C.  $} \\[0.8ex]
\hline\hline
\end{tabular*}

{\tiny Notes.

a. $H_0$ is given in Km/s/Mpc.

b. $N.C.$ means the results are not conclusive. The data do not constrain the parameters sufficiently.

}

\label{table:y1}
\end{table*}


\begin{table*}
\caption{{\small Table of best fits and their likelihoods (1$\sigma$) for redshift $y_4$, for the three sets of
         parameters $\mathcal{A}$, $\mathcal{B}$ and $\mathcal{C}$.
                 Set 1 of observations is Union 2 + HST. Set 2 of observations is Union 2 + HST + $H(z)$.}}

\begin{tabular*}{\textwidth}{@{\extracolsep{\fill}}ccccccc} 
\hline\hline

{\small Parameter}  &   {\small $\mathcal{A}$, Set 1}                  & {\small $\mathcal{A}$, Set 2 }
                    &   {\small $\mathcal{B}$, Set 1}                  & {\small $\mathcal{B}$, Set 2 }
                    &   {\small $\mathcal{C}$, Set 1}                  & {\small $\mathcal{C}$, Set 2 }  \\ [1.5ex]
                    
                    &   {\small $\chi^2_{min} = 530.3$ }                     & {\small 544.8 }
                    &   {\small 529.7 }                                      & {\small 544.6 }
                    &   {\small 529.9 }                                      & {\small 544.5 } \\[0.8ex]
\hline {\small $H_0$}      & {\small$74.55$}{\tiny${}_{-7.53}^{+7.54}$}      & {\small$73.71$}{\tiny ${}_{-5.24}^{+5.29}$}
                    & {\small $73.95$}{\tiny ${}_{-7.22}^{+7.99}$}           & {\small $73.43$}{\tiny ${}_{-5.74}^{+6.05}$}
                    & {\small $74.12$}{\tiny ${}_{-7.78}^{+8.27}$}           & {\small $73.27$}{\tiny ${}_{-5.91}^{+6.86}$} \\[0.8ex]

{\small $q_0$}      & {\small$-0.7492$}{\tiny${}_{-0.6228}^{+0.5899}$}      & {\small$-0.6504$}{\tiny ${}_{-0.3303}^{+0.4275}$}
                    & {\small $-0.4611$}{\tiny ${}_{-0.6710}^{+0.5422}$}     & {\small $-0.7230$}{\tiny ${}_{-0.4585}^{+0.5851}$}
                    & {\small $-0.4842$}{\tiny ${}_{-0.9280}^{+2.7126}$}     & {\small $-0.7284$}{\tiny ${}_{-0.4838}^{+0.6062}$} \\[0.8ex]

{\small $j_0$}      & {\small $2.558$}{\tiny${}_{-8.913}^{+7.441}$}     & {\small $1.342$}{\tiny ${}_{-1.780}^{+1.391}$}
                    & {\small $-3.381$}{\tiny ${}_{-2.149}^{+10.613}$}       & {\small $2.017$}{\tiny ${}_{-3.022}^{+3.149}$}
                    & {\small $-1.940$}{\tiny ${}_{-2.148}^{+8.041}$}        & {\small $2.148$}{\tiny ${}_{-4.036}^{+3.414}$} \\[0.8ex]

{\small $s_0$}      & {\small $9.85$}{\tiny${}_{-26.69}^{+74.69}$}  &{\small $3.151$}{\tiny ${}_{-1.771}^{+3.920}$}
                    & {\small $-37.67$}{\tiny ${}_{-60.10}^{+89.51}$}        & {\small $5.278$}{\tiny ${}_{-14.732}^{+13.076}$}
                    & {\small $-13.48$}{\tiny ${}_{-31.28}^{+71.65}$}        & {\small $2.179$}{\tiny ${}_{-35.919}^{+42.126}$} \\[0.8ex]

{\small $l_0$}      & {\Large --}                       & {\Large --} 
                    & {\small $N.C.$}                                        & {\small $-0.13$}{\tiny ${}_{-65.87}^{+96.75}$}
                    & {\small $N.C.$}                                        & {\small $-11.60$}{\tiny ${}_{-187.96}^{+193.88}$} \\[0.8ex]

{\small $m_0$}      & {\Large --}                        & {\Large --} 
                    & {\Large --}                                            & {\Large --} 
                    & {\small $N.C.$}                                        & {\small $70.9$}{\tiny ${}_{-2254.5}^{+2497.8}$} \\[0.8ex]

\hline\hline
\end{tabular*}

{\tiny Notes.

a. $H_0$ is given in Km/s/Mpc.

b. $N.C.$ means the results are not conclusive. The data do not constrain the parameters sufficiently.

}

\label{table:y4}
\end{table*}

\begin{figure}
\begin{center}
\includegraphics[width=2in]{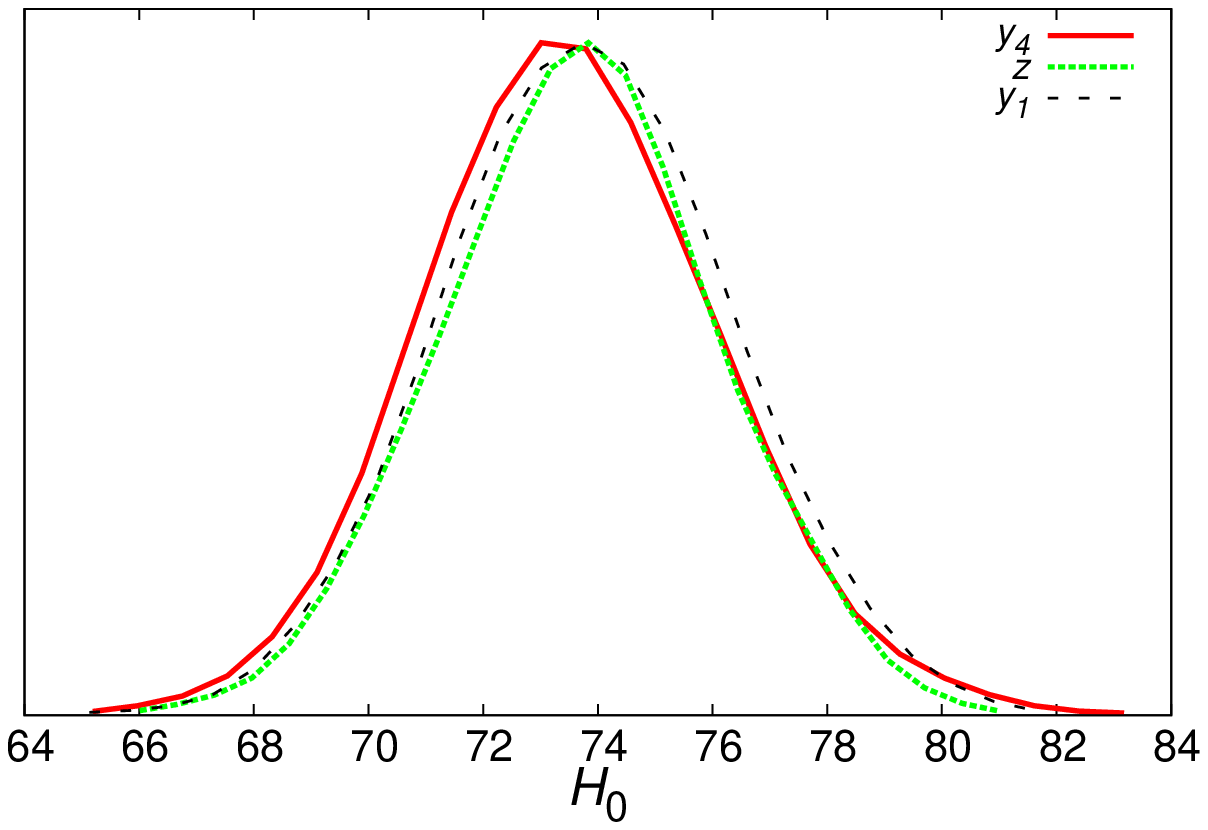}
\includegraphics[width=2in]{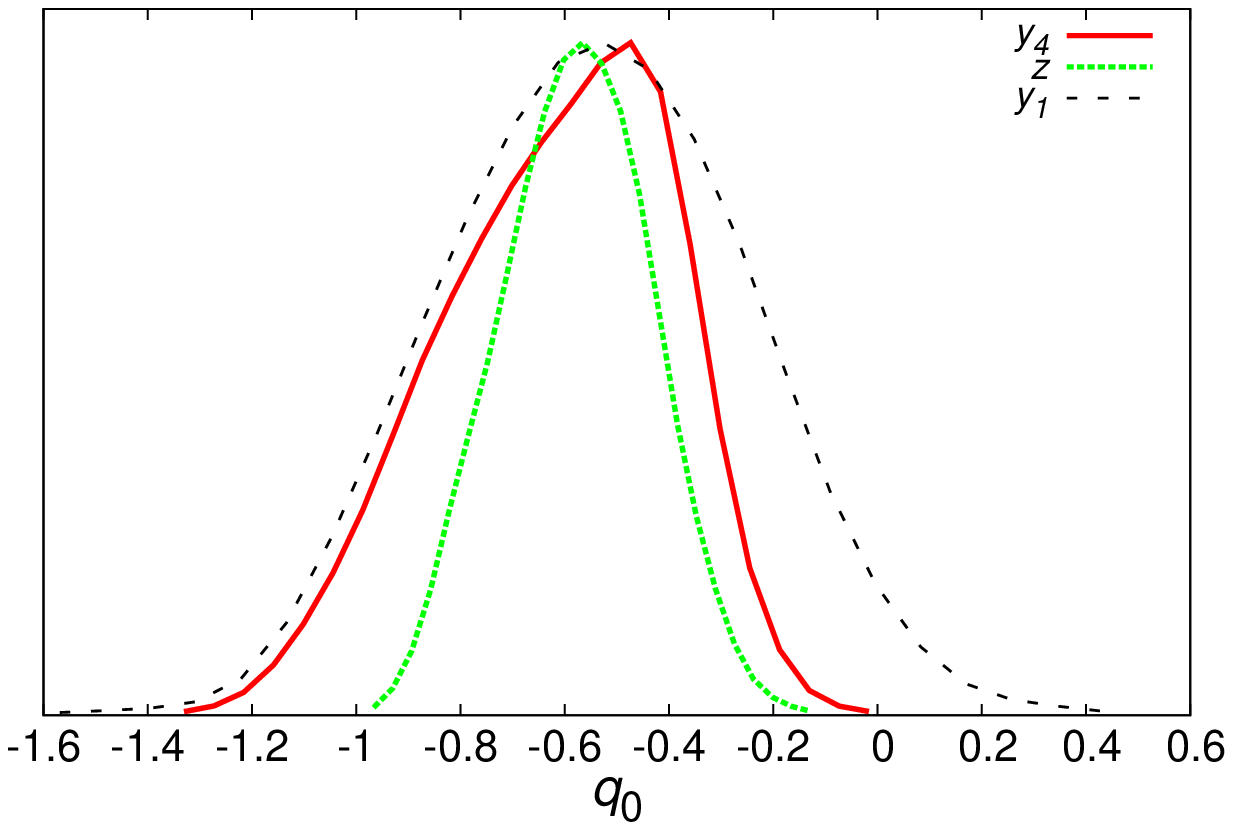}
\includegraphics[width=2in]{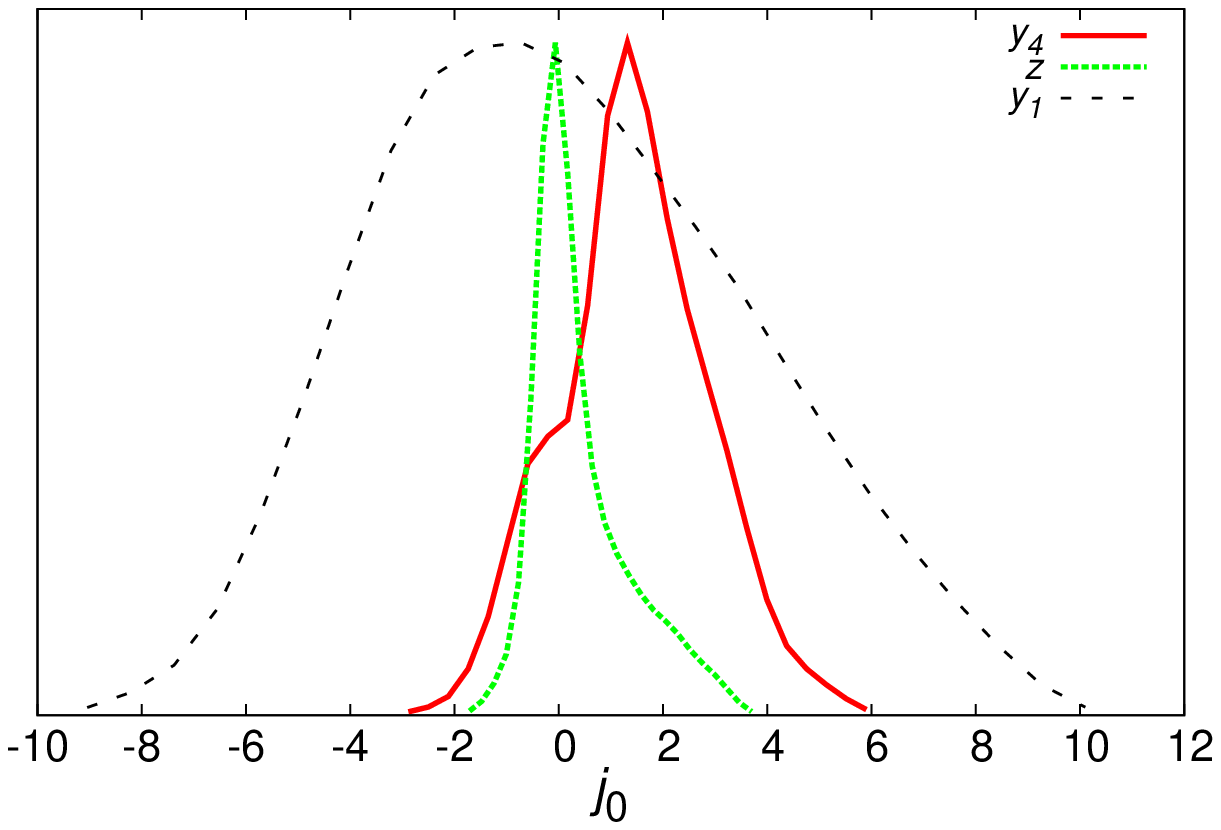}
\includegraphics[width=2in]{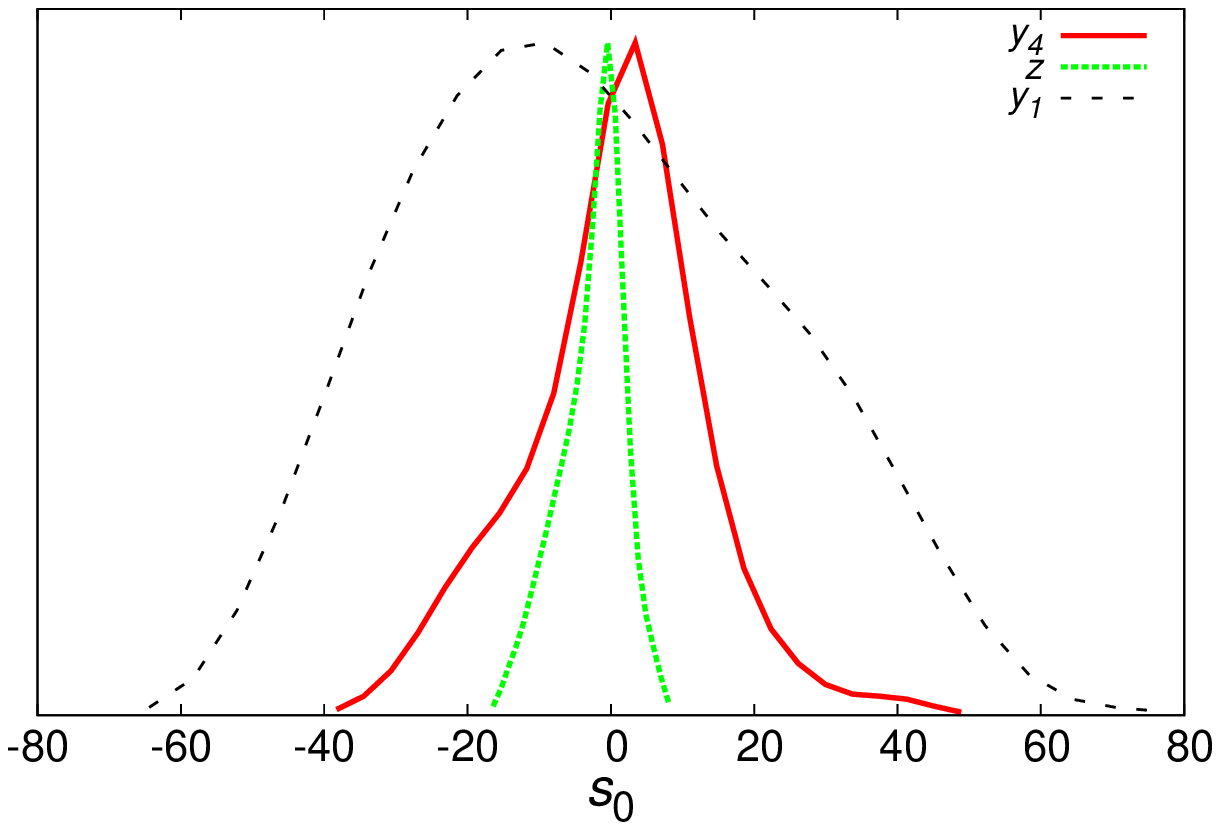}
\includegraphics[width=2in]{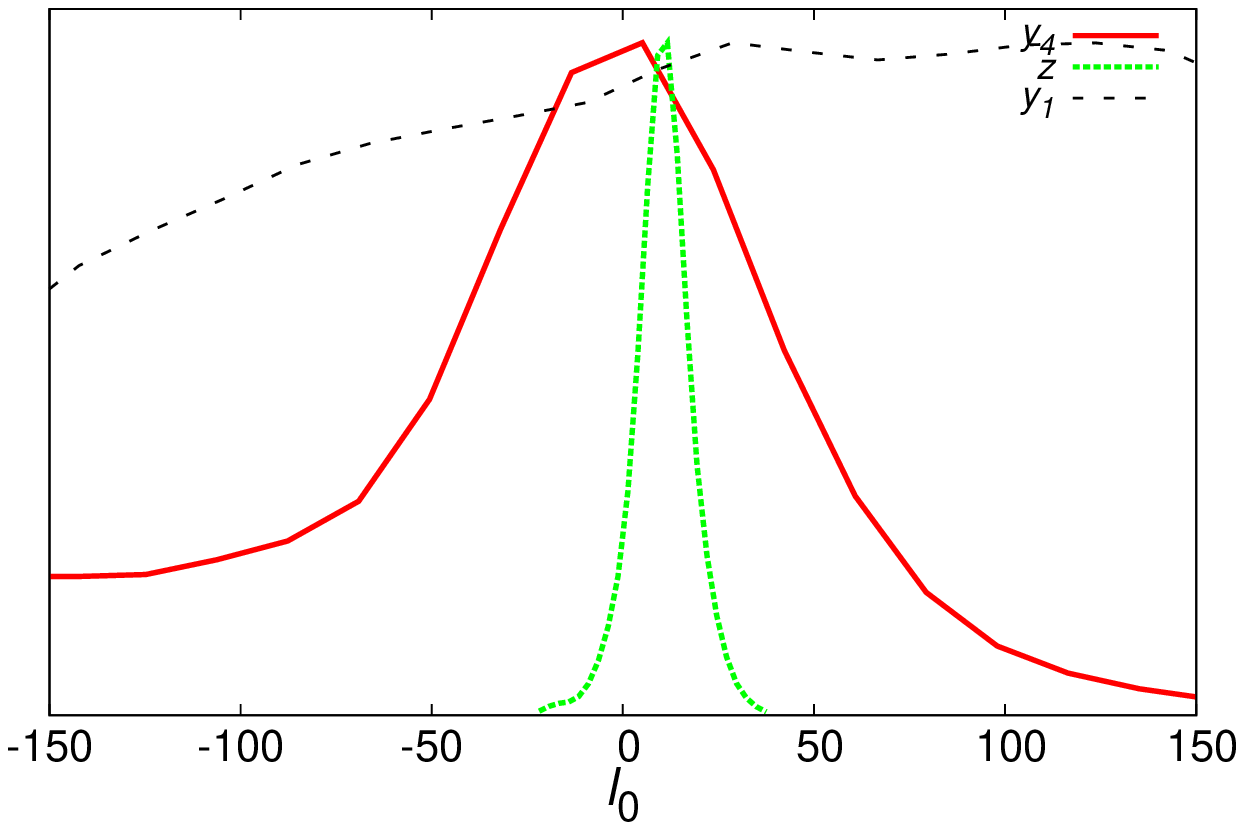}
\includegraphics[width=2in]{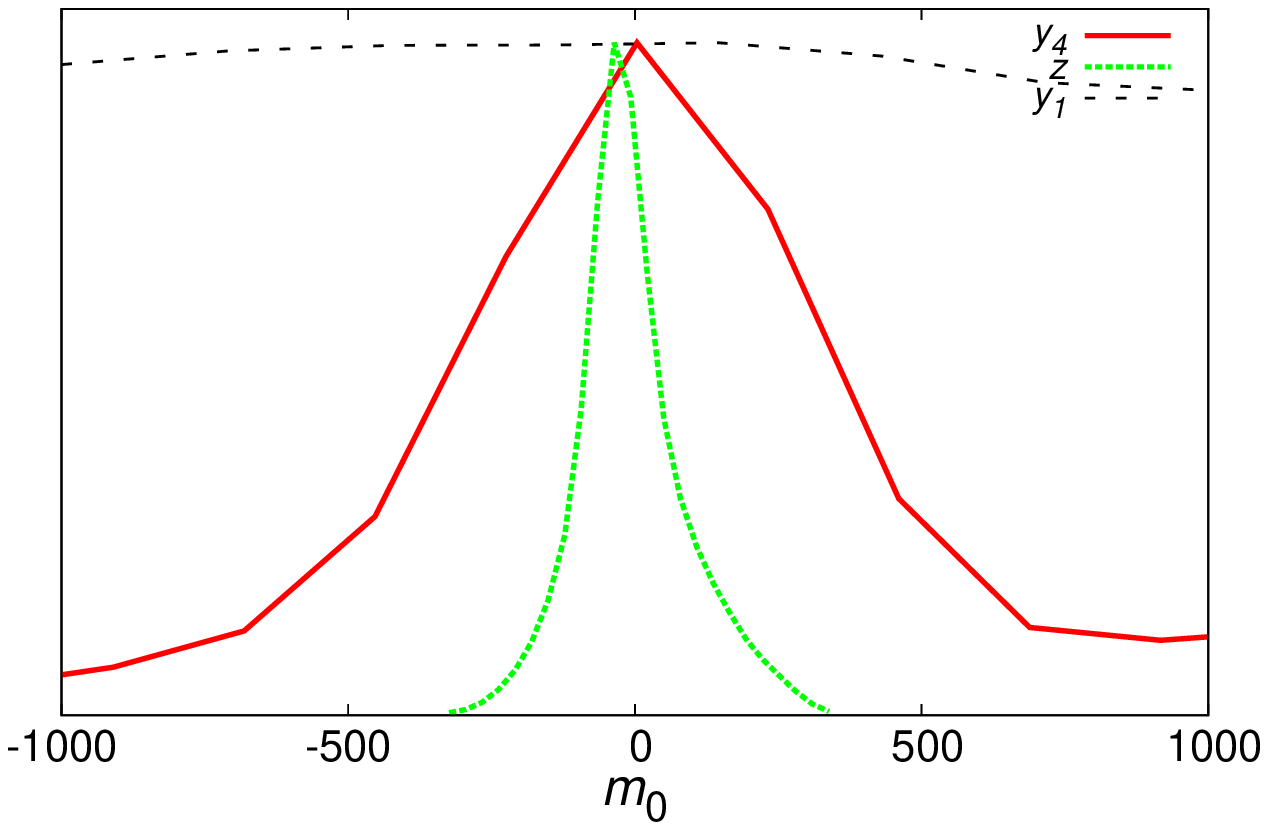}
{\small \caption{(color online) 1-dimensional marginalized posteriors for
the complete CS (parameter set $\mathcal{C}$), using set 2 of observations
(Union 2 + HST + $H(z)$). Dotted (green) line is redshift
$z$, dashed (black) line is $y_1$ and solid (red) line is $y_4$.}
\label{fig:1dim}}
\end{center}
\end{figure}

\begin{figure}
\begin{center}
\includegraphics[width=2in]{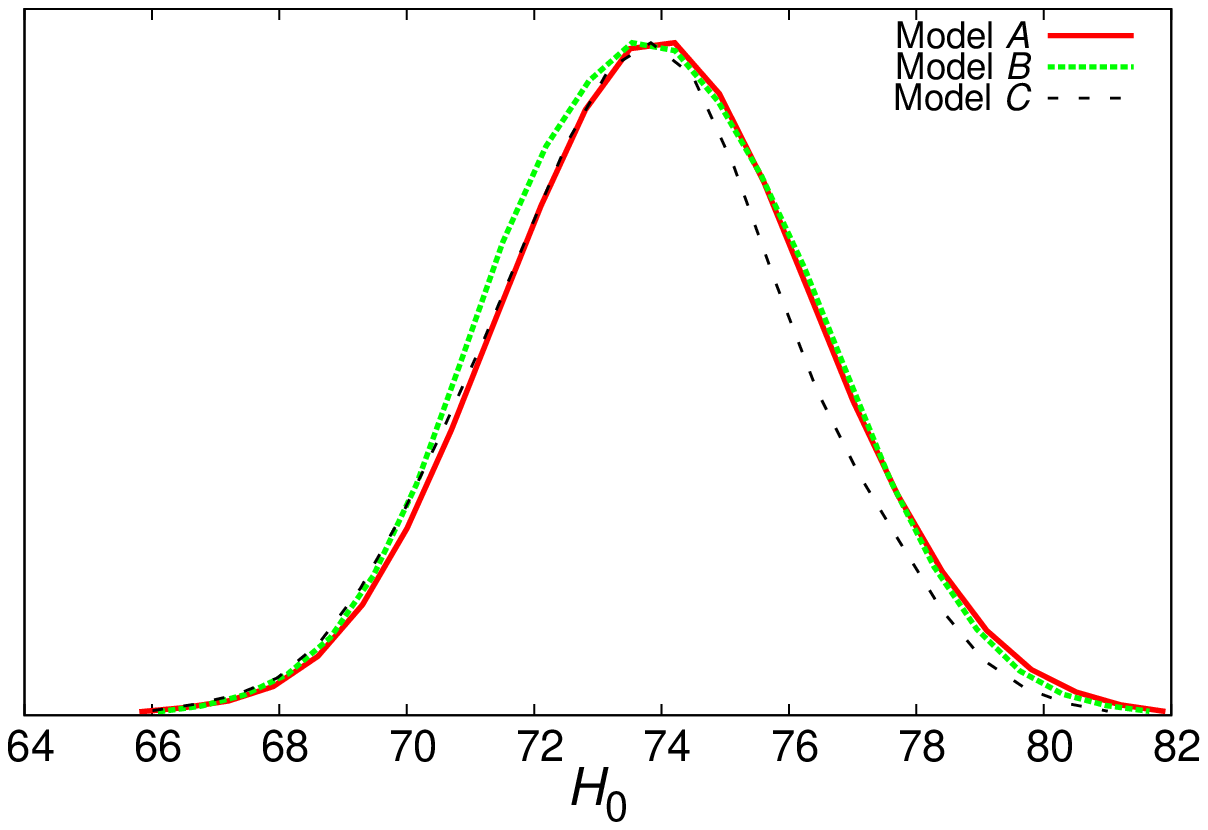}
\includegraphics[width=2in]{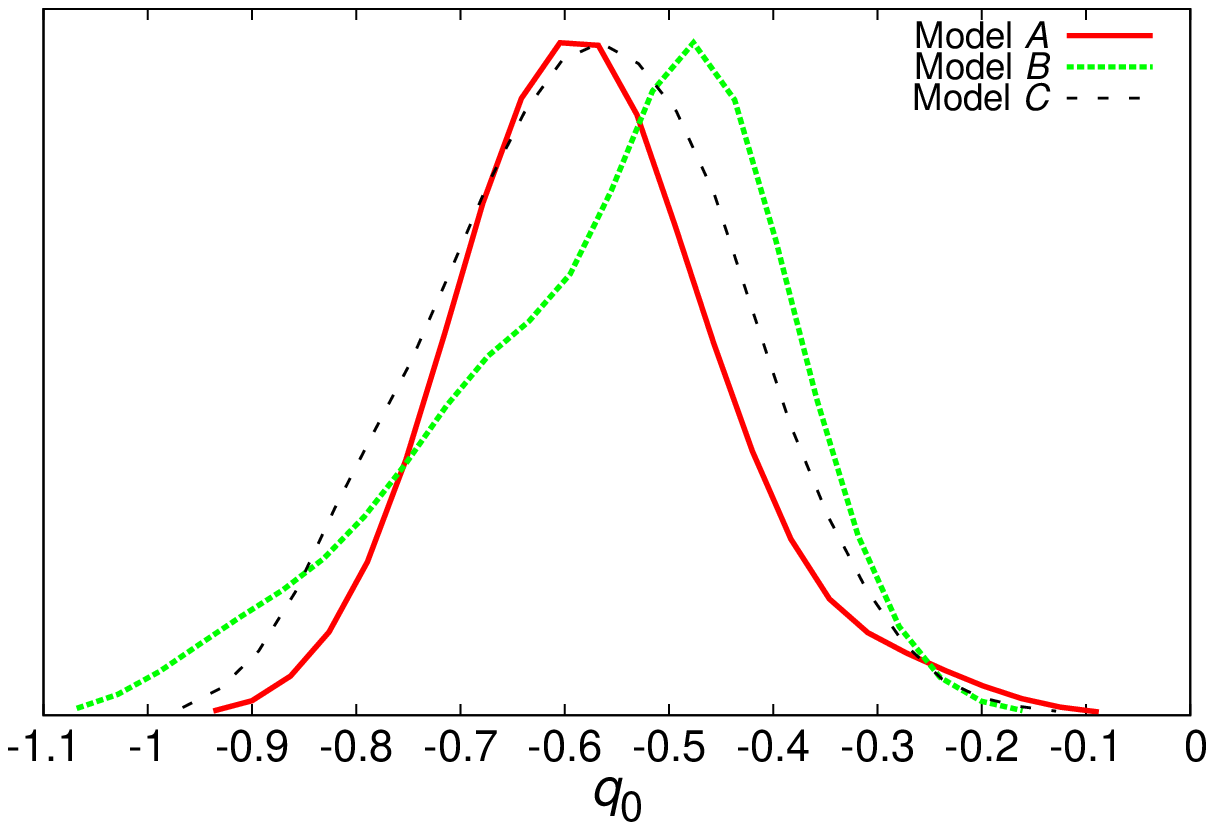}
\includegraphics[width=2in]{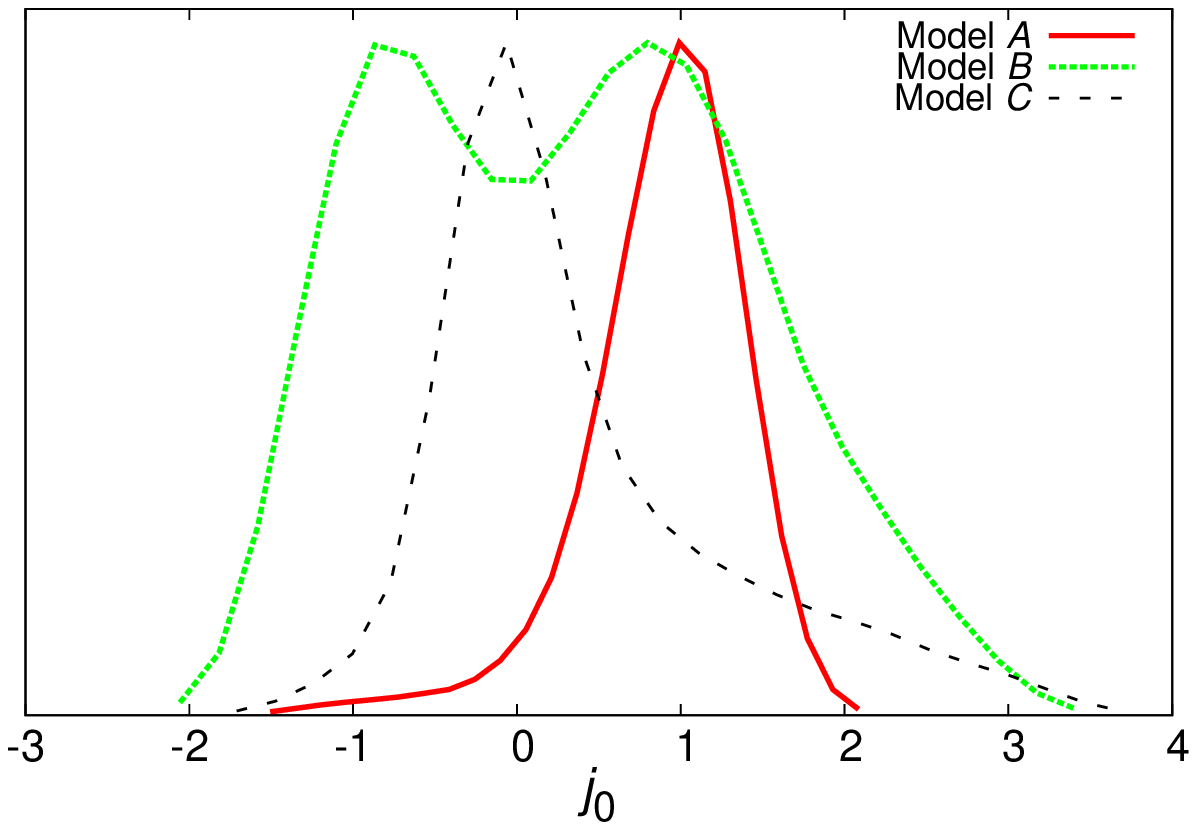}
\includegraphics[width=2in]{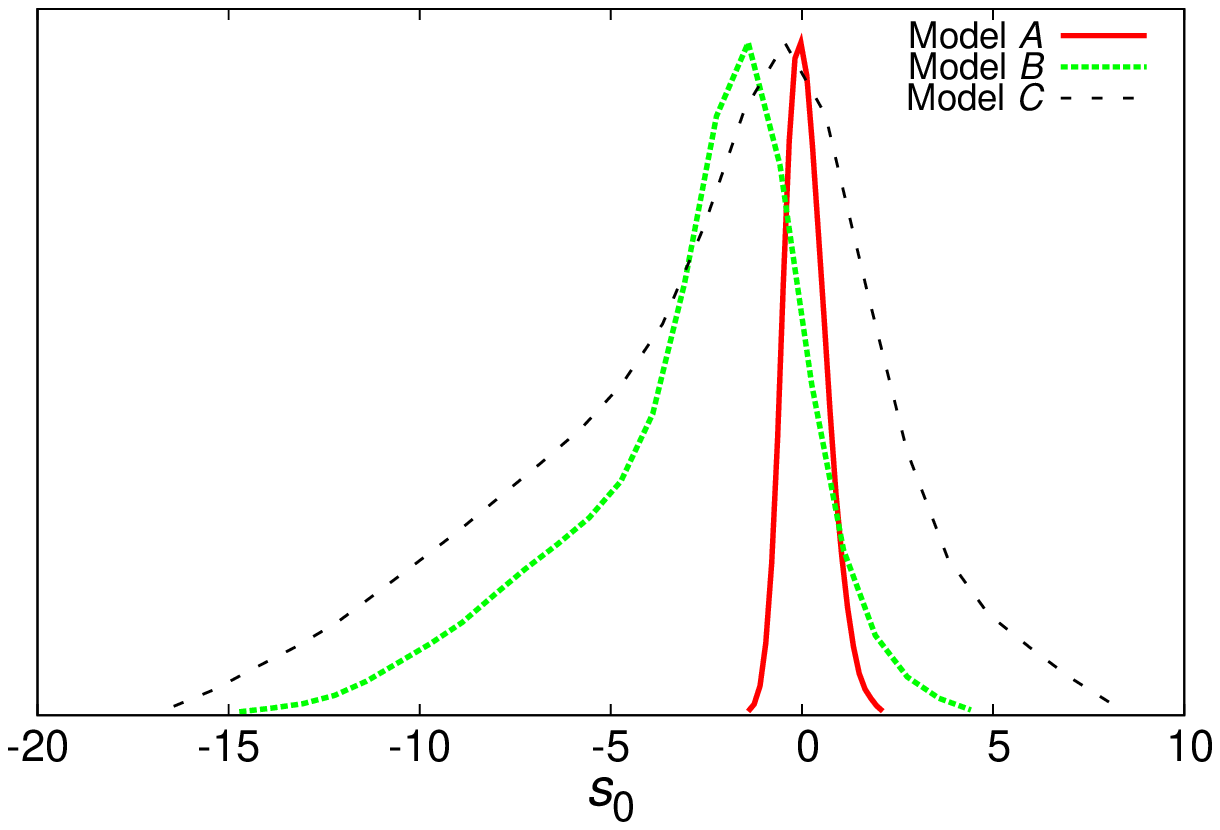}
{\small\caption{(color online) 1-dimensional marginalized posteriors for
$H_0$, $q_0$, $j_0$ and $s_0$, using set 2 of observations
(Union 2 + HST + $H(z)$). Solid (red) line is parameter set $\mathcal{A}$,
dotted (green) line is parameter set $\mathcal{B}$ and dashed (black) line is
parameter set $\mathcal{C}$.}
\label{fig:1dim2}}
\end{center}
\end{figure}

\begin{figure}
\begin{center}
\includegraphics[width=3.5 in]{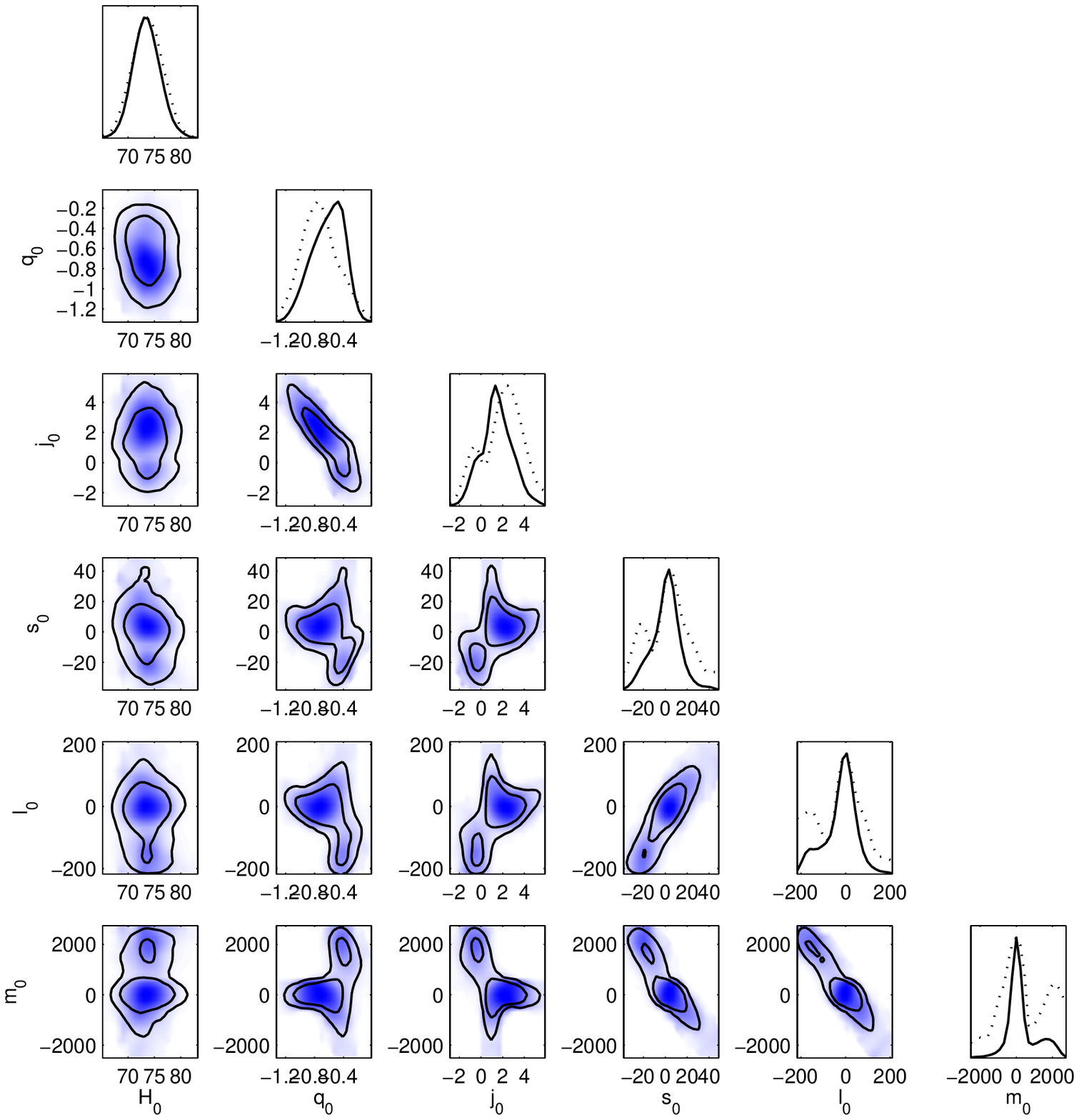}
{\small \caption{Marginalized posterior constraints for redshift
$y_4$ and parameter set $\mathcal{C}$, using set 2 of observations
(Union 2 + HST + $H(z)$). The shaded region and the dotted lines show
the likelihoods of the samples.}
\label{fig:2dim}}
\end{center}
\end{figure}

In Fig.~\ref{fig:1dim} we compare the 1-dimensional marginalized
posterior distributions for each parameter and each redshift for set
2 of observations. We note that the parameters $l_0$ and $m_0$ are
not well constrained when using the redshift $y_1$. By introducing
$y_4$ it is possible to overcome this issue, obtaining good results
on $l_0$ and $m_0$ as well. As expected, the redshift $z$ appears to
be statistically more favored than $y_4$. We confirm what we
conjectured in Sec.~II: the sixth order of the Hubble expansion in
$z$ works better than any other parametrization, if $z\leq1$.
Nonetheless, $y_4$ should be taken seriously as a possible
alternative to $y_1$, when $z\geq1$. Similar conclusions have been
drawn for the set 1; see Tables~\ref{table:z},~\ref{table:y1},
and~\ref{table:y4}. The only caveat is that for set 1 the
results appear to be less accurate as in the previous case.\\
We proceed to determine if the introduction of the
parameters $l_0$ and $m_0$ is convenient. To this end, in
Fig.~\ref{fig:1dim2} we plot the first four CS parameters' posterior
distributions for the three parameter sets, comparing their statistical
widths. We note that the dispersions are enlarged considerably when
we add the $l_0$ parameter; however, introducing $m_0$ does not
substantially broaden the posterior distributions; besides, the standard deviations
of the posteriors are in a proportion $\, 1:2.28:1.92 \,$ for $j_0$
and $\,1:5.66:8.38\,$ for $s_0$. \\
In Fig.~\ref{fig:2dim} we present
the summary of the results for the redshift $y_4$ and parameter set
$\mathcal{C}$ by plotting the 2-dimensional contours and the
likelihood samples.

\section{The connection between the CS and the EoS of the Universe}

In Sec.~I, we outlined that an expression for the EoS is naturally
associated to each cosmological fluid in a given cosmological
model; in particular, in thermodynamics the EoS
characterizes the properties of such a fluid.
Under the assumption of a specific gravitational model, the quest of understanding the
expansion history of the Universe is equivalent to reproducing the
correct EoS during different phases. Many physical mechanisms are
hidden in the EoS parameter $\omega$. Finding the correct EoS has
thus high importance in cosmology, because it offers
a key to understanding the micro-physics associated to DE and/or DM.
In this section we will assume that General Relativity gives a correct
description of gravity at the scales under consideration.

Under the hypotheses of cosmography, we cannot \emph{a priori}
assume an EoS of the Universe, because we were not specifying any
particular cosmological model at the beginning. We recall that the
EoS of the Universe is given by
$\omega = \sum_iP_i/ \sum_i\rho_i$, where the subindex $i$ refers to
the different fluids that the Universe comprises. Hence, to evaluate
$\omega$, one needs to know the total pressure,
$P =\sum_iP_i$, and the total density,
$\rho = \sum_i\rho_i$. Even though we do not assume any EoS
explicitly, it is possible to expand the pressure in terms of the
cosmic time or redshift variables, i.e. $z,y_1,y_4$. By expanding
the pressure into a series, it is possible to predict the values of
its derivatives with respect to the cosmic time or the redshift
variables. In fact, one can relate the derivatives of $P$ to the CS;
therefore, by substituting the values of CS in terms of these
derivatives into the luminosity distances we are able to directly
fit the parameters of the EoS of the Universe from luminosity
distance data.

The reason of constraining the pressure derivatives lies in the
possibility of discriminating among models; in principle, a
model which does not satisfy such bounds can be easily
discarded. The expansion of $P$ in terms of the cosmic time is
formally given by
\begin{equation}
P = \sum_{k=0}^{\infty}\frac{1\,}{k!}\frac{d^{k}P}{dt^{k}}\Big|_{t_0}(t-t_0)^k
=\sum_{k=0}^{\infty}\frac{1\,}{k!} \frac{d^{k}P}{dy_i^{k}}\Big|_{0}y_i^k\,,
\end{equation}
where $y_i=z,\,y_1,\,y_4$. By truncating the series at the fourth
order, and the continuity equation
\begin{equation}\label{continuità}
\frac{d\rho}{dt}+3H(P+\rho)=0\,,
\end{equation}
and the Friedmann equation $H^2=\frac{1}{3}\rho$, we can write down an explicit dependence of the pressure and
the coefficients
$\frac{d^kP}{dt^k}$ on the CS as follows\footnote{Here
in  Eqs.~($\ref{eq:pressureandD}$) we are considering
$\frac{8\pi G}{3}=c=1$ for brevity. These factors are
considered again in the numerical simulations.}

\begin{subequations} \label{eq:pressureandD}
\begin{align}
 P &= \frac{1}{3}H^2 \left( 2q - 1\right)\,,\label{eq:pressure} \\
\frac{dP}{dt} & = \frac{2}{3} H^3 \left(1 - j\right)\,, \label{eq:pressure1} \\
\frac{d^2P}{dt^2} & = \frac{2}{3} H^4 \left(j- 3 q - s -3 \right)\,,
    \label{eq:pressure2} \\
\frac{d^3P}{dt^3} & = \frac{2}{3} H^5 \Bigl[(2s + j - l + q\,( 21 -
j) +
    6 q^2 + 12\Bigr]\,, \label{eq:pressure3}\\
\frac{d^4P}{dt^4} & = \frac{2}{3} H^6 \Bigl[j^2 +  3 l -m
    - 144 q - 81 q^2\\ &- 6 q^3 -12 j\left(2 + q\right) - 3 s - 3 q\,
    s-60\Bigr]\,,
    \label{eq:pressure4}
\end{align}
\end{subequations}
where we evaluated the derivatives up to the order of $m_0$.\\
We list the coefficients $\frac{d^kP}{dz^k}$, $\frac{d^kP}{dy_1^k}$ and
$\frac{d^kP}{dy_4^k}$ in Appendix~\ref{app:pressure}.
For completeness, we write down the
transformation laws between $z,y_1,y_4$ and the cosmic time $t$, i.e.
\begin{eqnarray}\label{eq:derivativeconversions}
 \frac{\partial}{\partial t} &\rightarrow& -(1+z) H \cdot \frac{\partial}{\partial z} \nonumber\\
  &\rightarrow& -(1+y_1) H \frac{\partial}{\partial y_1} \\
  &\rightarrow& - \cos y_4 (\cos y_4 + \sin y_4) H \frac{\partial}{\partial y_4}\,. \nonumber
\end{eqnarray}
In addition, by using Eq.~($\ref{continuità}$),
Eqs.~\eqref{eq:pressureandD} and the Friedmann equation $H^2=\frac{1}{3}\rho$, we find the expression
for the EoS parameter of the Universe as
\begin{equation}\label{eq:omega}
\omega = \frac{2q-1}{3}\,.
\end{equation}

\subsection{Fitting the EoS}

In this subsection, our goal is to obtain constraints on the EoS and the pressure
derivatives by inverting Eqs. \eqref{eq:pressure1} -
\eqref{eq:pressure4} and by rewriting the luminosity distance as a
function of
$\omega_0,\frac{dP}{dy_i}\Big|_{0},\frac{d^2P}{dy_i^2}\Big|_{0},\frac{d^3P}{dy_i^3}\Big|_{0},\frac{d^4P}{dy_i^4}\Big|_{0}$.\\
In other words, we use Eqs.~\eqref{eq:pressure1} -
\eqref{eq:pressure4} and~\eqref{eq:omega}, indicated henceforth by
\begin{equation*}
 \mathcal{D} = \left\{\omega,P_1:=\frac{dP}{dy_i}, P_2:=\frac{d^2P}{dy_i^2},
    P_3:=\frac{d^3P}{dy_i^3},P_4:=\frac{d^4P}{dy_i^4}\right\},
\end{equation*}
to express the vector $\mathcal{CS}\equiv \{q_0,j_0,s_0,l_0,m_0\}$
as a function of the EoS parameter and the pressure derivatives,
$\mathcal{CS}=\mathcal{CS}(\omega_0,\,P_1,\,P_2,\,P_3,\,P_4)$. The
purpose is to plug those results into the expressions for the
luminosity distance in order to find numerical best fit values of
the new set of parameters, using $d_L=d_L(\omega_0,P_1,P_2,P_3,P_4)$
in the numerical analysis.

In principle, there exists also an alternative procedure, which
consists in taking the results already obtained for the CS and to
propagate the errors through
Eqs.~\eqref{eq:pressure1}-\eqref{eq:pressure4}, without performing
another fitting procedure.
But in choosing this way, we would face an unacceptable
increase in the errors, as opposed to the direct fit of
$\{\omega_0,P_1,P_2,P_3,P_4\}$. Thus, in order to reduce the error
propagation, the simplest and most straightforward way is to evaluate the
coefficients by a direct fit of the luminosity distance.
The search for the best-fit values for the new set of parameters
is performed by using the procedure of MCMC simulations developed in Sec.~III.


For statistical reasons, we choose set 2 of observations, which is more
complete and suitable for this kind of fit.

The explicit expressions for the luminosity distance in terms of the different
redshift parameters are reported for completeness in
Appendix~\ref{app:EoS}. As in Sec.~III we limit our analysis to
$d_L$.

\begin{figure*}
\begin{center}
\includegraphics[width=2in]{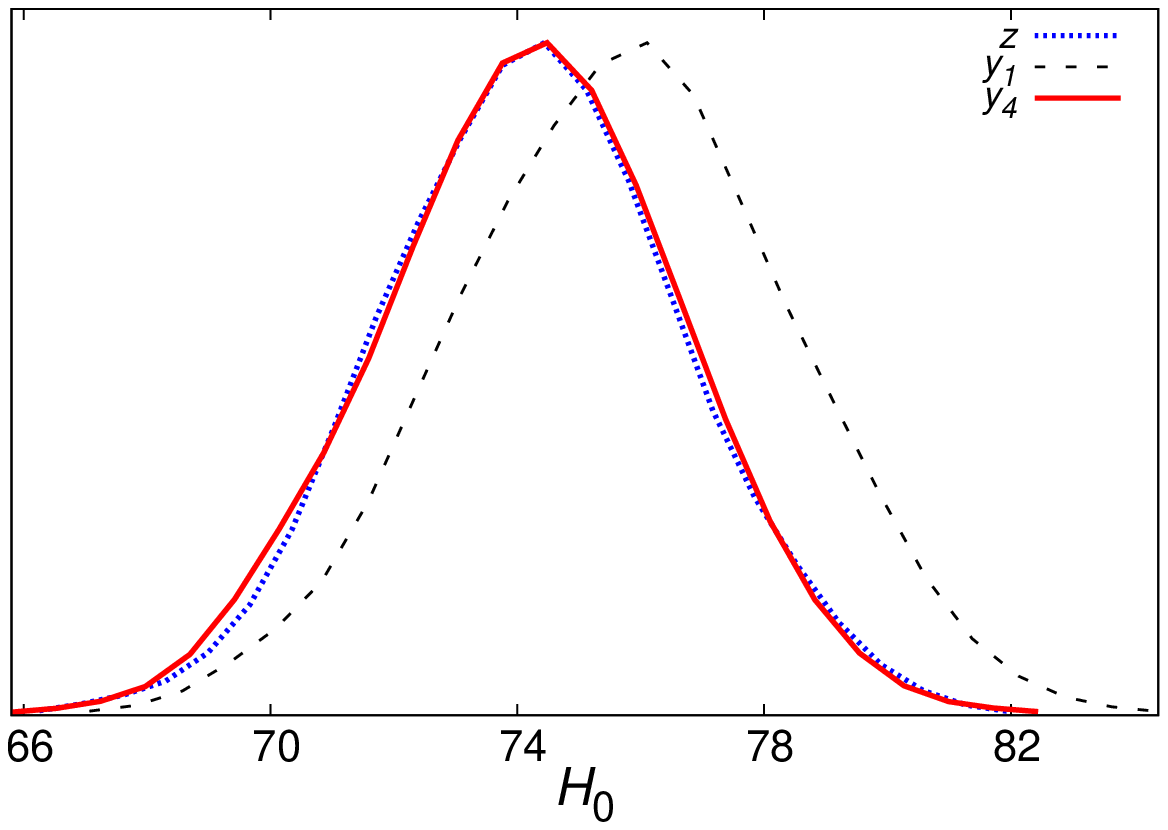}
\includegraphics[width=2in]{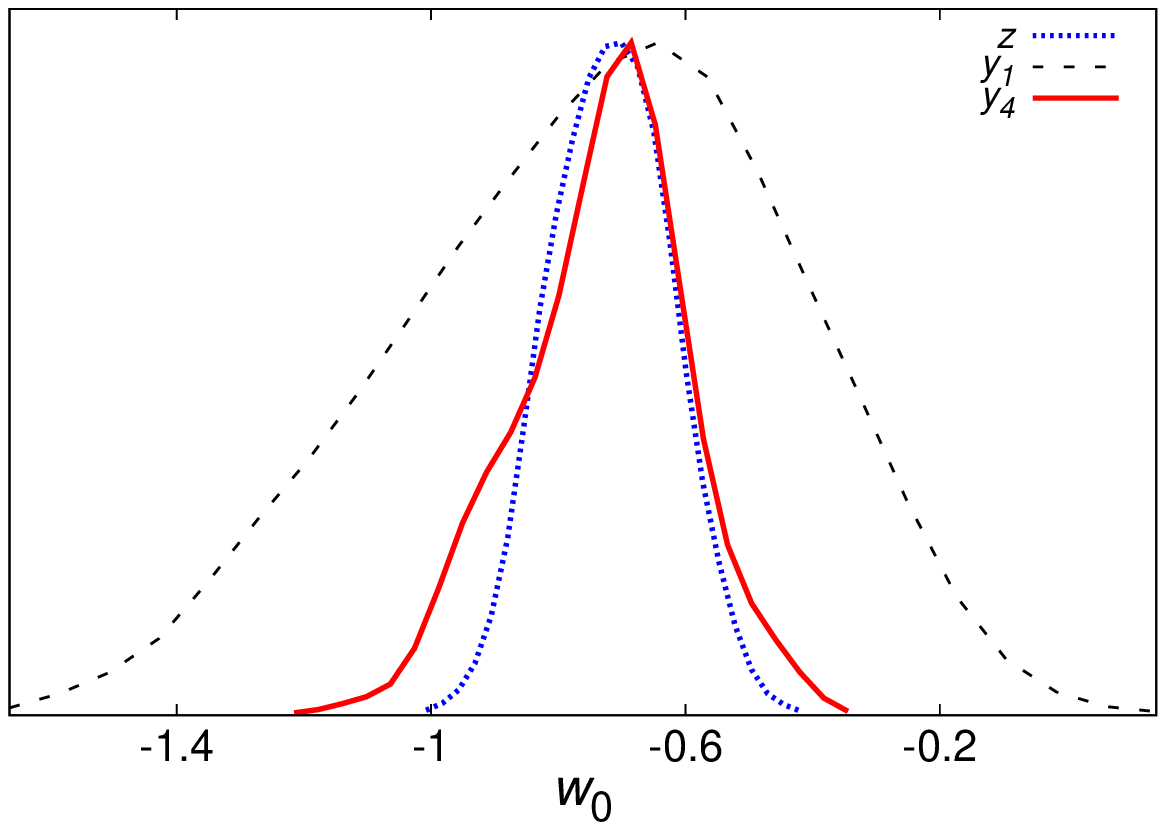}
\includegraphics[width=2in]{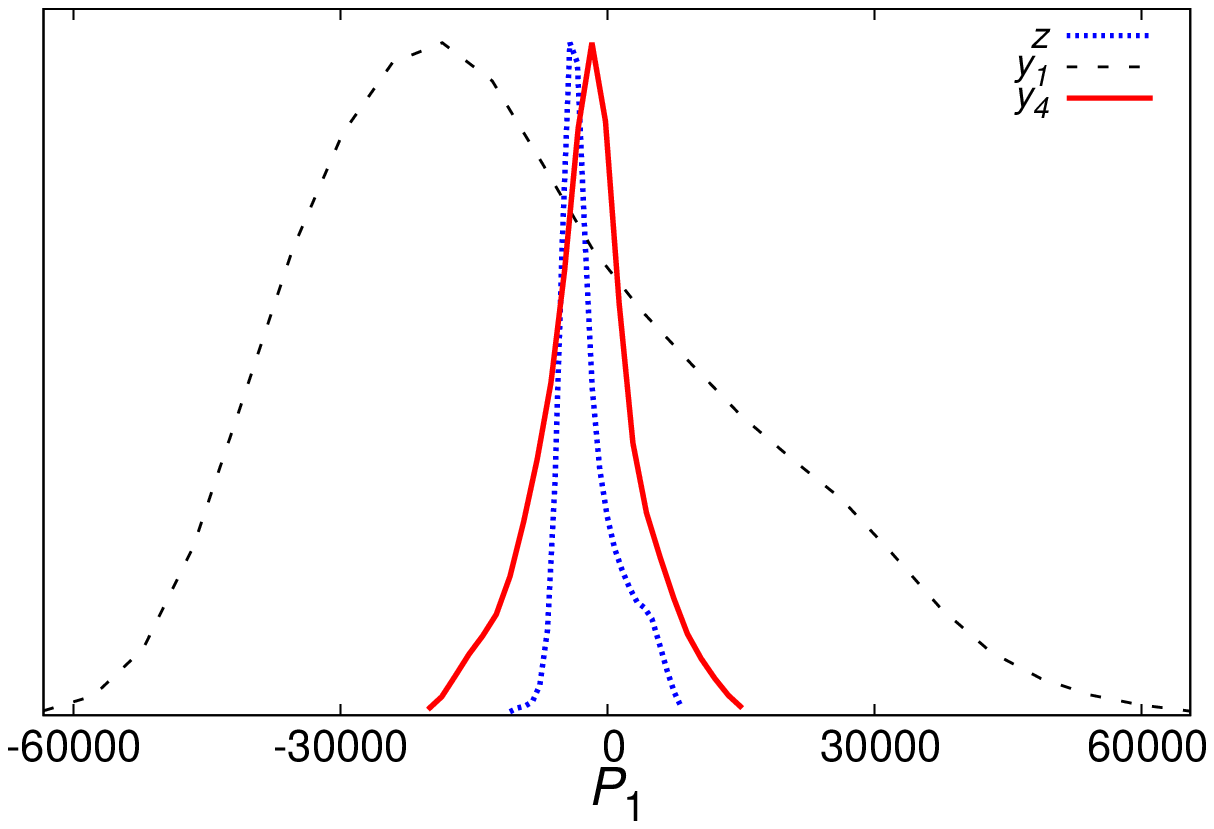}
\includegraphics[width=2in]{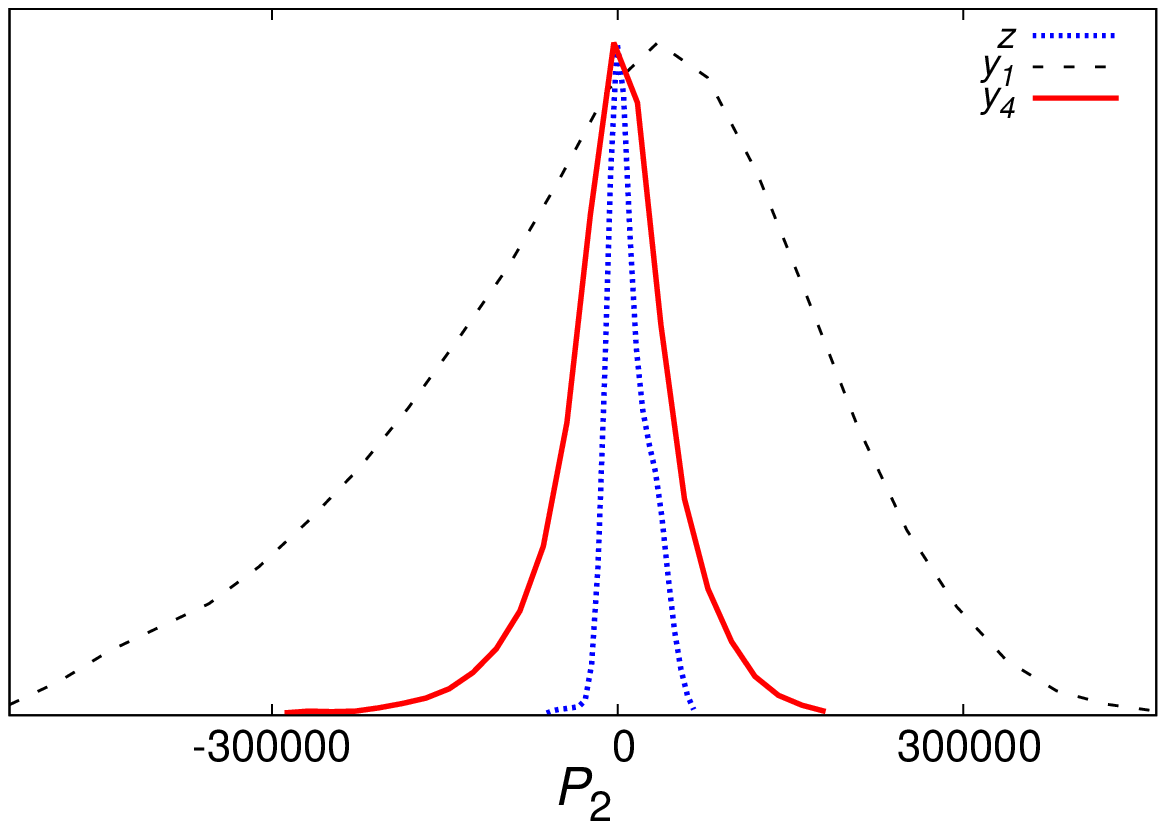}
\includegraphics[width=2in]{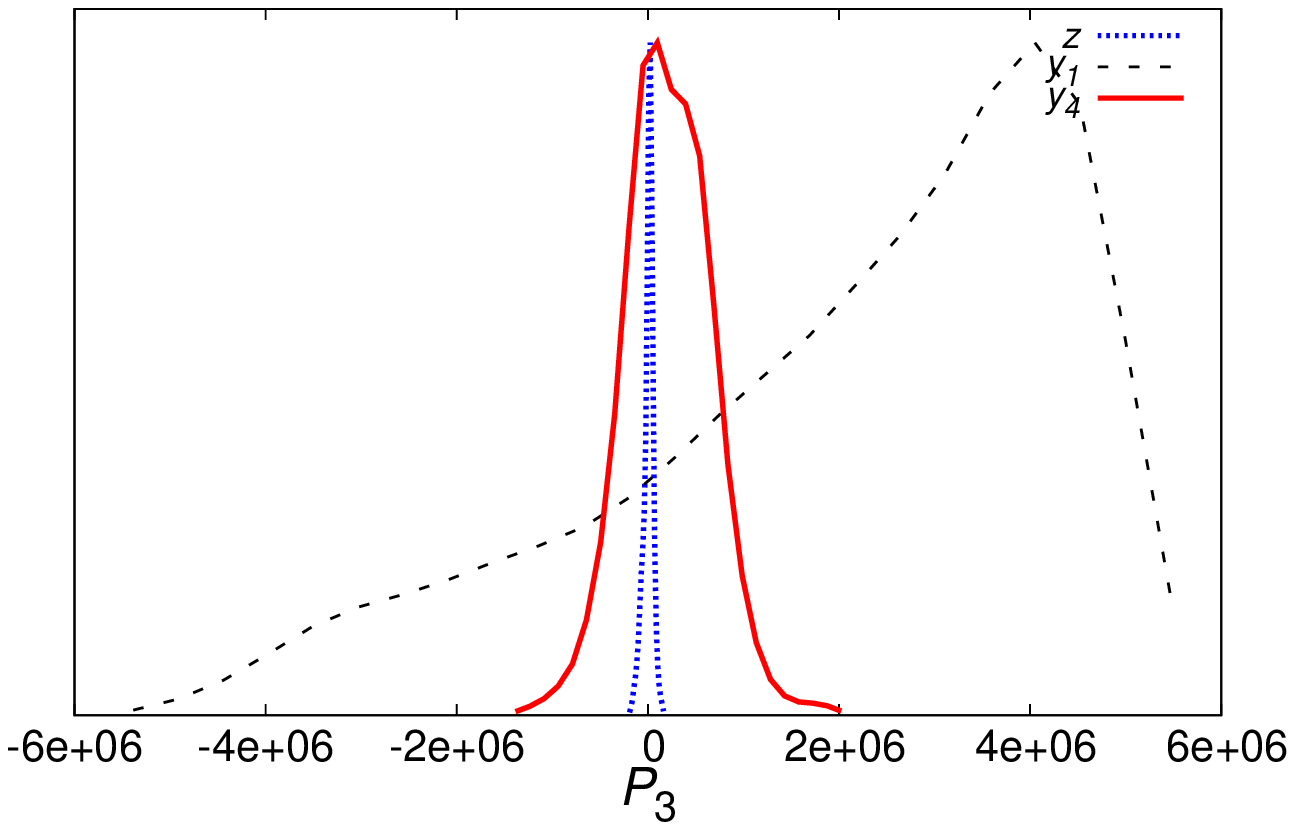}
\includegraphics[width=2in]{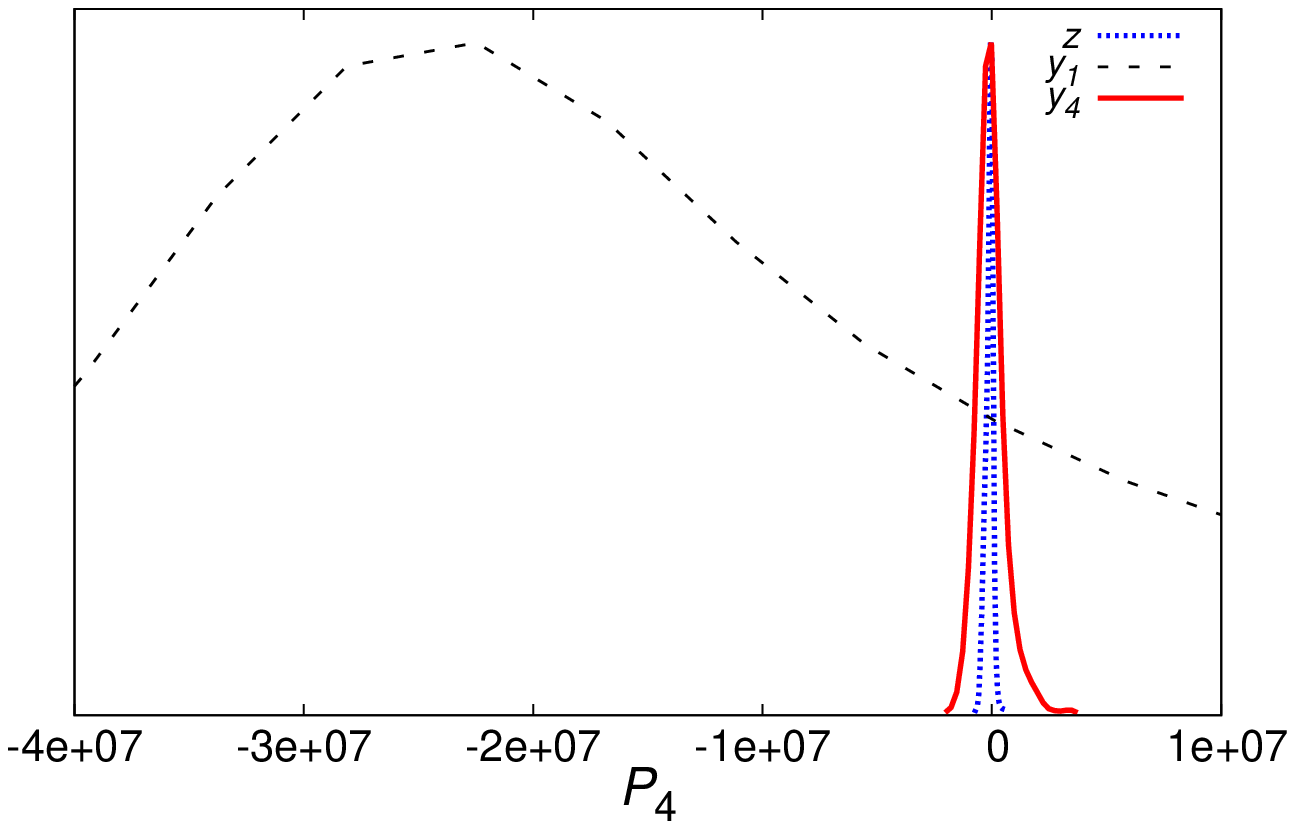}
{\small \caption{(color online) 1-dimensional marginalized posteriors for
the complete set of parameters of the EoS analysis, using set 2 of
observations (Union 2 + HST + $H(z)$).
Dotted (blue) lines are used for $z$, dashed (black) lines for $y_1$, and
solid (red) lines are for redshift $y_4$.}
\label{Fig:EoS}}
\end{center}
\end{figure*}

Figure~\ref{Fig:EoS} shows the obtained marginalized posteriors and
in Tab.~\ref{table:EoS} we  present the summary of the results. We
observe the same hierarchy of redshifts as in analysis of Sec. III; meaning that
our new ``redshift'' introduction, $y_4$, appears
to be statistically favored with respect to $y_1$.

\begin{table*}
\caption{{\small Table of mean values of the posteriors and their likelihoods (1$\sigma$) for the three
          redshifts, using set 2 of observations (Union 2 + HST + $H(z)$). $P_1(z)$, $P_1(y_1)$,
          $P_1(y_4)$, $P_2(z)$, $P_2(y_4)$ and $P_3(z)$ are in units of $10^4\, c^2/\kappa$,
          $P_2(y_1)$ and $P_4(y_4)$ in units of $10^5\, c^2/\kappa$, $P_3(y_4)$ and
          $P_4(z)$ in units of $10^6\, c^2/\kappa$, $P_3(y_1)$ in units of $10^7\,
          \frac{c^2}{\kappa}$, and $P_4(y_1)$ in units of $10^8\, c^2/\kappa$.}}

\begin{tabular*}{0.75\textwidth}{@{\extracolsep{\fill}}cccc}
\hline\hline

$\quad$ {\small Parameter}  $\quad$ & {\small Redshift $z$}
                    & {\small Redshift $y_1$}
                    & {\small Redshift $y_4$}  \\ [1.5ex]

\hline {\small$H_0$}       & {\small
$74.23$}{\tiny${}_{-2.36}^{+2.31}$}
                    & {\small $74.20$}{\tiny ${}_{-2.36}^{+2.37}$}
                    & {\small $75.70$}{\tiny ${}_{-2.66}^{+2.68}$} \\[0.8ex]

{\small $\omega_0$}      & {\small
$-0.7174$}{\tiny${}_{-0.0964}^{+0.0922}$}
                    & {\small $-0.7439$}{\tiny ${}_{-0.3222}^{+0.3085}$}
                    & {\small $-0.7315$}{\tiny ${}_{-0.1373}^{+0.1193}$} \\[0.8ex]

{\small $P_1$}      & $\quad$ {\small
$-0.209$}{\tiny${}_{-0.261}^{+0.347}$}
$\quad$
                    & $\quad$ {\small $-0.991$}{\tiny ${}_{-2.213}^{+2.393}$} $\quad$
                    & $\quad$ {\small $-0.228$}{\tiny${}_{-0.528}^{+0.506}$} $\quad$\\[0.8ex]

{\small $P_2$}      & {\small $0.988$}{\tiny${}_{-1.539}^{+2.012}$}
                    & {\small $-0.134$}{\tiny ${}_{-1.729}^{+1.623}$}
                    & {\small $-0.246$}{\tiny${}_{-3.927}^{+4.133}$} \\[0.8ex]

{\small $P_3$}      & {\small $0.630$}{\tiny${}_{-4.932}^{+4.010}$}
                    & {\small $0.205$}{\tiny${}_{-0.257}^{+0.294}$}
                    & {\small $0.217$}{\tiny${}_{-0.400}^{+0.625}$} \\[0.8ex]

{\small $P_4$}      & {\small $-0.107$}{\tiny${}_{-0.170}^{+0.099}$}
                    & {\small $-0.150$}{\tiny${}_{-0.187}^{+0.209}$}
                    & {\small $-0.289$}{\tiny ${}_{-6.112}^{+4.690}$} \\[0.8ex]

\hline \hline

\end{tabular*}

{\tiny Notes.

a. $H_0$ is given in Km/s/Mpc.

}
\label{table:EoS}
\end{table*}

\subsection{Comparison with models}

The results found by fitting the EoS show no conclusive evidence for a pressure
varying in time, as shown by the $1\sigma$ confidence levels in Tab.~ \ref{table:EoS}.
This means that a negative constant pressure model is favored for
depicting the cosmic speed up. Thus, the only two models accounting our results appear to be the concordance model and the vanishing speed of sound model (VSSM), proposed in \cite{Luongo2011,nobel}.
Their EoS parameters (neglecting radiation components at late times) are $\omega=-1/(1+ \Omega_m/\Omega_{\Lambda} a^{-3})$ and $\omega=-1/(1-(\xi-\Omega_m/\Omega_{X}) a^{-3})$, for $\Lambda$CDM and VSSM respectively. They imply $w_0 \simeq - 0.73$ nowadays, by using the values of $\Omega_m\simeq0.27$ and $\Omega_X\simeq0.78$, $\xi\simeq-0.025$. These results have been confirmed by the present analysis.

Using set 2 of observations we worked out the
$\Lambda$CDM model, which is for our purposes and the redshifts
involved sufficiently described by the two parameters $\{\Omega_m h^2,\theta\}$.
The philosophy is to estimate $\Omega_m h^2$ and $\theta$ by the
Monte Carlo simulation and then substitute it into the CS in the
$\Lambda$CDM model. To evaluate the CS for $\Lambda$CDM we use Eqs.~
($\ref{eq:CSoftime}$) with $H=H_0\sqrt{\Omega_m(1+z)^3+1-\Omega_m}$, yielding

\begin{eqnarray}
q_0 &=&  -1 + \frac{3}{2}\Omega_m\,, \nonumber\\
j_0  &=&  1\,, \nonumber\\
s_0 &=&  1 - \frac{9}{2} \Omega_m\,, \\
l_0 &=& 1 +  3 \Omega_m - \frac{27}{2}\Omega_m^2\,, \nonumber\\
m_0 &=&  1 - \frac{27}{2}\Omega_m^2 - 81\Omega_m^2 -
\frac{81}{2}\Omega_m^3\,. \nonumber
\end{eqnarray}
Any significative tension between the CS values obtained in this way
and the values derived before using cosmography would be an
indication of the validity of a different theory, other than the
concordance model. Table~\ref{table:LCDM} gives the summary of the
likelihoods of the estimated and derived parameters. Comparing these
values with the ones obtained for models $\mathcal{A}$,
$\mathcal{B}$ and $\mathcal{C}$ we note that all our results are
compatible with the
$\Lambda$CDM model within the limits of error. \\

Now, we want to factorize the effects of DE by assuming a cold dark matter
model, for which the Friedmann equation is given by

\begin{equation*}
 H=H_0\sqrt{\Omega_m(1+z)^3+G(z)}\,,
\end{equation*}
where $G(z)$ models the corresponding DE term.
We want to obtain constraints on $G(z)$ and its derivatives with respect
to redshifts for the present time $z=0$. To this purpose we calculate
the derivatives of the Hubble parameter as stated above with respect
to the redshift and equate the results to the derivatives of the Hubble
parameter in terms of the CS. Thus we obtained the derivatives of $G(z)$
with respect to $z$ (results see Appendix~\ref{app:G}). The same
has been done for the redshifts $y_1$ and $y_4$.

We can now use the numerically obtained values for the CS and
the value $\Omega_m=0.274{\tiny{}_{-0.015}^{+0.015}}$ obtained by
the WMAP7 collaboration~\cite{WMAP7}
to calculate the $G^{(k)}(y_i)$ and, by error propagation,
their $1\sigma$ error bars. The results can be found in Table~\ref{tab:G}.
With these values it is possible to investigate the compatibility of
a model of dark energy characterized by the function $G(z)$ with the
observational data at present time. The value of $G(z)$ evaluated today can
be estimated by the flat space condition, which implies $G_0 = G(z=0) = 1- \Omega_m$ or,
assuming Gaussian distributions, $G_0 = 0.726{\tiny{}_{-0.015}^{+0.015}}$.

From  Table~\ref{tab:G}, we conclude that our results are consistent with
a constant function $G(z) = \Omega_{\Lambda}$, which is the case of the $\Lambda$CDM model, or the case of an emergent constant as shown in \cite{Luongo2011,nobel}.

\begin{table}
\caption{{\small Table of best fits and their likelihoods (1$\sigma$) for the
estimated (top panel) and derived (lower panel) parameters
for the $\Lambda$CDM model, using set 2 of observations (union 2 + HST + $H(z)$).}}

\begin{tabular}{cc}
\hline\hline 
$\qquad$ Parameter $\qquad$  &    $\qquad$ Best Fit $\,$ (1$\sigma$)$\qquad$                \\[0.8ex]
\hline
 $\Omega_m h^2$              & {\small $0.1447$}{\tiny ${}_{-0.0174}^{+0.0181}$}         \\[0.8ex]
 $\theta$                    & {\small $1.060 $}{\tiny ${}_{-0.022 }^{+0.020 }$}         \\[0.8ex]
\hline
 $H_0$                       & {\small $74.05$}{\tiny ${}_{-7.19}^{+7.90}$}         \\[0.8ex]

 $q_0$                       & {\small $-0.6633$}{\tiny ${}_{-0.6580}^{+0.5753}$}   \\[0.8ex]

 $j_0$                       & {\small $1$}                                         \\[0.8ex]

 $s_0$                       & {\small $-0.2061$}{\tiny${}_{-0.2015}^{+0.1772}$}          \\[0.8ex]

 $l_0$                       & {\small $2.774$}{\tiny${}_{-0.382}^{+0.485}$}          \\[0.8ex]

 $m_0$                       & {\small $-8.827$}{\tiny${}_{-2.941}^{+2.263}$}          \\[0.8ex]
\hline\hline
\end{tabular}

{\tiny Notes.

 $H_0$ is given in Km/s/Mpc. Here $h$ is defined through the relation $H_0 = 100\,h \text{km/s/Mpc}$, and $\theta$ is the ratio of the sound horizon to the angular diameter distance at recombination.

} 
\label{table:LCDM}
\end{table}

\begin{table*}
\caption{{\small Table of derived values and their likelihoods
(1$\sigma$) for the derivatives of $G(y_i)$ for the three redshifts
$z,y_1,y_4$, evaluated at $t=t_0$, using set $\mathcal{C}$ of
parameters and set 2 of observations (union 2 + HST + $H(z)$).}}

\begin{tabular}{cccc}
\hline\hline
 ~~Parameter~~  &    $\qquad$ Best Fit for $z\,$ (1$\sigma$)$\qquad$  &   $\qquad$ Best Fit for $y_1\,$ (1$\sigma$)$\qquad$ &  $\qquad$ Best Fit for $y_4\,$ (1$\sigma$)$\qquad$  \\[0.8ex]
\hline
 $G^{(1)}_0$      & {\small $0.0068$}{\tiny ${}_{-1.23}^{+1.299}$} & {\small $0.25$}{\tiny ${}_{-1.87}^{+1.57}$}& {\small $-0.2788$}{\tiny ${}_{-1.43882}^{+1.59994}$}        \\[0.8ex]

 $G^{(2)}_0$  & {\small $-2.22$}{\tiny ${}_{-3.33}^{+4.61}$} & {\small $-4.71$}{\tiny ${}_{-8.8}^{+9.13}$}& {\small $1.7384$}{\tiny ${}_{-4.94215}^{+4.88827}$}  \\[0.8ex]

 $G^{(3)}_0$  & {\small $13.65$}{\tiny ${}_{-5.8}^{+8.04}$}  & {\small $0.7476$}{\tiny ${}_{-49.34}^{+55.73}$}& {\small $-3.43039$}{\tiny ${}_{-12.4275}^{+13.3403}$}                                       \\[0.8ex]

 $G^{(4)}_0$  & {\small $-17.24$}{\tiny${}_{-31.89}^{+35.27}$} &  {\small $N.C.$}  & {\small $-16.2906$}{\tiny ${}_{-47.602}^{+50.8794}$}    \\[0.8ex]

 $G^{(5)}_0$  & {\small $-129.74$}{\tiny${}_{-130.39}^{+174.29}$}   &  {\small $N.C.$}  & {\small $-8.74593$}{\tiny ${}_{-323.676}^{+332.408}$}       \\[0.8ex]

\hline\hline
\end{tabular}

{\tiny Notes.

a. $N.C.$ means the results are not conclusive. The data do not constrain the parameters sufficiently.

}

\label{tab:G}
\end{table*}

\section{Conclusions}

In this work we emphasized the importance of constraining the
Universe dynamics through the use of a model-independent procedure,
which does not \emph{a priori} assume the validity of a
particular cosmological model.
In other words, we used the so called cosmography, sometimes
also referred to as cosmokinetics, to investigate the kinematics
of the Universe. By following~\cite{CattViss2008}, we
performed an analysis combining theoretical derivations of cosmological
distances and numerical data fitting, using the Union 2 compilation,
together with the HST and the $H(z)$ samples. For the fitting, we
introduced new parameterizations in addition to the conventional
redshift $z$ to express the cosmological distances, underlining their importance to
avoid divergences at high redshifts and to increase the accuracy of the
analysis. Moreover, we proposed prescriptions to build up new viable
parametrizations, able to overcome these issues.

We considered three further parameterizations, predicting
that only one is really a viable option for fitting. By using the
luminosity distance, i.e. $d_L$, in terms of the standard redshift $z$,
of the alternative parametrization $y_1\equiv\frac{z}{1+z}$ and of
the newly introduced $y_4=\arctan(z)$, we obtained bounds
on the cosmographic series. We moreover showed that there was no physical
reason to use other notions of cosmological distances than the luminosity
distance for evaluating bounds to the CS, as instead previously
reported in~\cite{CattViss2008}. The reason for this lies in the fact
that $d_L$ is adapted the best to the cosmological
data under consideration. Then, we also showed that the most successful
parametrization, apart from $z$, is represented by
$y_4$, as theoretically predicted in Sec. II.

We carried out our fits up to the sixth order in the CS, introducing a
further cosmographic parameter, namely $m_0$. In addition, we showed
that fitting $m_0$ together with $q_0,j_0,s_0$ and $l_0$, is quite
feasible to improve the accuracy of the analysis by fixing more stringent
limits on the CS, as opposed to the expectation that an additional
fitting parameter would significantly broaden the posterior distributions.

The analysis was done for three different sets of parameters
(including parameters of the CS with different maximum order) and
two different sets of observational data. We essentially found that
our numerical results appear to be fairly well in agreement with a
constant pressure associated to the fluid driving the cosmic
acceleration. At first sight, this fluid could be obviously thought
to originate from a cosmological constant, as depicted in the
standard concordance model. Meanwhile, our constraints would be
able, in future developments, to discard different classes of
models, which do not satisfy the numerical bounds.

In addition to the numerical fits for the CS, we proposed a way to
constrain both the EoS of the Universe as a whole and the derivatives of
pressure, by fitting them directly from the luminosity distance, up
to the order of the $m_0$ coefficient. In other words, we rewrote
$d_L$ in terms of the EoS parameter and the pressure derivatives and
performed another MCMC analysis. To achieve this, it is necessary to depart from model independent cosmography by assuming a specific gravitational theory. We choose General Relativity as valid, and make use of the Friedmann equations. The corresponding results
lead to a set of pressure derivatives compatible with zero within
the $1\sigma$ error propagation. Moreover, the EoS parameter $\omega$
is compatible with the one predicted by $\Lambda$CDM,i.e.
$\omega = -1/(1+ \Omega_m/\Omega_{\Lambda} a^{-3})$, obtaining
$\omega_0 \simeq - 0.73$. Even through the pressure derivative results seemed to
confirm $\Lambda$CDM, they also left open the possibility that the
viable model does not necessarily feature a cosmological constant,
but dark energy with constant pressure and a varying barotropic
factor, i.e. the VSSM model (as proposed in~\cite{Luongo2011,nobel}). In this model,
which does not involve a cosmological constant, the predicted bounds
are in agreement with our fitting results. Unfortunately, the strong
degeneracy with $\Lambda$CDM in fitting data leaves open the
question of which model is theoretically the favored one.

Future perspectives in this direction include using more
accurate datasets and constraining the analysis at higher redshifts,
in order to obtain limits able to discriminate among the two
paradigms. We expect that this could give further insight into
the issue of inferring the correct cosmological paradigm,
employing a model-independent procedure, which does not need to specify
a model a priori.

\acknowledgements

A.A. acknowledges  CONACYT for grant no. 215819.
C.G. is supported by the Erasmus Mundus Joint Doctorate Program by
Grant Number 2010-1816 from the EACEA of the European Commission.
O.L. wants to thank Prof. A. Melchiorri for useful discussions.

\begin{widetext}
\appendix

\section{5 Distances in terms of redshifts}
\subsection{Redshift $z$}
\label{app:z}
We begin with the results for the cosmological distances for the conventional redshift $z$,
starting with luminosity distance,
\begin{eqnarray*}
    d_L & = & \frac{1}{H_0} \cdot \Bigl[ z + z^2 \cdot \Bigl(\frac{1}{2} - \frac{q_0}{2} \Bigr) +
    z^3 \cdot \Bigl(-\frac{1}{6} -\frac{j_0}{6} + \frac{q_0}{6} + \frac{q_0^2}{2} \Bigr) + \nonumber\\
    &&+\, z^4 \cdot \Bigl( \frac{1}{12} + \frac{5 j_0}{24} - \frac{q_0}{12} + \frac{5 j_0 q_0}{12} -
    \frac{5 q_0^2}{8} - \frac{5 q_0^3}{8} + \frac{s_0}{24} \Bigr) + \nonumber\\
    &&+\, z^5 \cdot \Bigl( -\frac{1}{20} - \frac{9 j_0}{40} + \frac{j_0^2}{12} - \frac{l_0}{120} +
    \frac{q_0}{20} - \frac{11 j_0 q_0}{12} + \frac{27 q_0^2}{40} - \frac{7 j_0 q_0^2}{8} + \frac{11 q_0^3}{8} +
    \frac{7 q_0^4}{8} - \frac{11 s_0}{120} - \frac{q_0 s_0}{8} \Bigr) + \nonumber\\
    &&+\, z^6 \cdot \Bigl( \frac{1}{30} + \frac{7 j_0}{30} - \frac{19 j_0^2}{72} + \frac{19 l_0}{720} +
    \frac{m_0}{720} - \frac{q_0}{30} + \frac{13 j_0 q_0}{9} - \frac{7 j_0^2 q_0}{18} + \frac{7 l_0 q_0}{240}
    - \frac{7 q_0^2}{10} + \frac{133 j_0 q_0^2}{48} - \frac{13 q_0^3}{6} + \nonumber\\
    &&+\, \frac{7 j_0 q_0^3}{4} - \frac{133 q_0^4}{48} - \frac{21 q_0^5}{16} + \frac{13 s_0}{90}
    - \frac{7 j_0 s_0}{144} + \frac{19 q_0 s_0}{48} + \frac{7 q_0^2 s_0}{24} \Bigr) \Bigr]\,,
\end{eqnarray*}
the photon flux distance,
\begin{eqnarray*}
    d_F & = & \frac{1}{H_0} \cdot \Bigl[ z - z^2 \cdot \frac{q_0}{2} + z^3 \cdot \Bigl(-\frac{1}{24}
    -\frac{j_0}{6} + \frac{5q_0}{12} + \frac{q_0^2}{2} \Bigr)+ \nonumber\\
    &&+\, z^4 \cdot \Bigl( \frac{1}{24} + \frac{7 j_0}{24} - \frac{17 q_0}{48} + \frac{5 j_0 q_0}{12}
    - \frac{7 q_0^2}{8} - \frac{5 q_0^3}{8} + \frac{s_0}{24} \Bigr)  + \nonumber\\
    &&+\, z^5 \cdot \Bigl( -\frac{71}{1920} - \frac{47 j_0}{120} + \frac{j_0^2}{12} - \frac{l_0}{120}
    + \frac{149 q_0}{480} - \frac{9 j_0 q_0}{8} + \frac{47 q_0^2}{40} - \frac{7 j_0 q_0^2}{8}
    + \frac{27 q_0^3}{16} + \frac{7 q_0^4}{8} - \frac{9 s_0}{80} - \frac{q_0 s_0}{8} \Bigr)  + \nonumber\\
    &&+\, z^6 \cdot \Bigl( \frac{31}{960} + \frac{457 j_0}{960} - \frac{11 j_0^2}{36} + \frac{11 l_0}{360}
    + \frac{m_0}{720} - \frac{1069 q_0}{3840} + \frac{593 j_0 q_0}{288} - \frac{7 j_0^2 q_0}{18}
    + \frac{7 l_0 q_0}{240} - \frac{457 q_0^2}{320} + \frac{77 j_0 q_0^2}{24} + \nonumber\\
    &&-\, \frac{593 q_0^3}{192} + \frac{7 j_0 q_0^3}{4} - \frac{77 q_0^4}{24} - \frac{21 q_0^5}{16}
    + \frac{593 s_0}{2880} - \frac{7 j_0 s_0}{144} + \frac{11 q_0 s_0}{24} + \frac{7 q_0^2 s_0}{24} \Bigr) \Bigr]\,,
\end{eqnarray*}
the photon count distance,
\begin{eqnarray*}
    d_P & = & \frac{1}{H_0} \cdot \Bigl[ z +z^2 \cdot \Bigl( -\frac{1}{2} - \frac{q_0}{2} \Bigr)
    + z^3 \cdot \Bigl( \frac{1}{3} -\frac{j_0}{6} + \frac{2 q_0}{3} + \frac{q_0^2}{2} \Bigr) + \nonumber\\
    &&+\, z^4 \cdot \Bigl( -\frac{1}{4} + \frac{3 j_0}{8} - \frac{3 q_0}{4} + \frac{5 j_0 q_0}{12}
    - \frac{9 q_0^2}{8} - \frac{5 q_0^3}{8} + \frac{s_0}{24} \Bigr)  + \nonumber\\
    &&+\, z^5 \cdot \Bigl( \frac{1}{5} - \frac{3 j_0}{5} + \frac{j_0^2}{12} - \frac{l_0}{120}
    + \frac{4 q_0}{5} - \frac{4 j_0 q_0}{3} + \frac{9 q_0^2}{5} - \frac{7 j_0 q_0^2}{8} + 2 q_0^3
    + \frac{7 q_0^4}{8} - \frac{2 s_0}{15} - \frac{q_0 s_0}{8} \Bigr)  + \nonumber\\
    &&+\, z^6 \cdot \Bigl( -\frac{1}{6} + \frac{5 j_0}{6} - \frac{25 j_0^2}{72} + \frac{5 l_0}{ 144}
    + \frac{m_0}{720} - \frac{5 q_0}{6} + \frac{25 j_0 q_0}{9} - \frac{7 j_0^2 q_0}{18}
    + \frac{7 l_0 q_0}{240} - \frac{5 q_0^2}{2} + \frac{175 j_0 q_0^2}{48} - \frac{25 q_0^3}{6} + \nonumber\\
    &&+\, \frac{7 j_0 q_0^3}{4} - \frac{175 q_0^4}{48} - \frac{21 q_0^5}{16} + \frac{5 s_0}{18}
    - \frac{7 j_0 s_0}{144} + \frac{25 q_0 s_0}{48} + \frac{7 q_0^2 s_0}{24} \Bigr) \Bigr]\,,
\end{eqnarray*}
the deceleration distance,
\begin{eqnarray*}
    d_Q & = & \frac{1}{H_0} \cdot \Bigl[ z + z^2 \cdot \Bigl( -1 - \frac{q_0}{2} \Bigr)
    + z^3 \cdot \Bigl( \frac{23}{24} -\frac{j_0}{6} + \frac{11 q_0}{12} + \frac{q_0^2}{2} \Bigr) + \nonumber\\
    &&+\, z^4 \cdot \Bigl( -\frac{11}{12} + \frac{11 j_0}{24} - \frac{61 q_0}{48} + \frac{5 j_0 q_0}{12}
    - \frac{11 q_0^2}{8} - \frac{5 q_0^3}{8} + \frac{s_0}{24} \Bigr)  + \nonumber\\
    &&+\, z^5 \cdot \Bigl( \frac{563}{640} - \frac{17 j_0}{20} + \frac{j_0^2}{12} - \frac{l_0}{120}
    + \frac{253 q_0}{160} - \frac{37 j_0 q_0}{24} + \frac{51 q_0^2}{20} - \frac{7 j_0 q_0^2}{8}
    + \frac{37 q_0^3}{16} + \frac{7 q_0^4}{8} - \frac{37 s_0}{240} - \frac{q_0 s_0}{8} \Bigr)  + \nonumber\\
    &&+\, z^6 \cdot \Bigl( -\frac{1627}{1920} + \frac{1273 j_0}{960} - \frac{7 j_0^2}{18}
    + \frac{7 l_0}{180} + \frac{m_0}{720} - \frac{7141 q_0}{3840} + \frac{1037 j_0 q_0}{288}
    - \frac{7 j_0^2 q_0}{18} + \frac{7 l_0 q_0}{240} - \frac{1273 q_0^2}{320} + \nonumber\\
    &&+\, \frac{49 j_0 q_0^2}{12} - \frac{1037 q_0^3}{192} + \frac{7 j_0 q_0^3}{4}
    - \frac{49 q_0^4}{12} - \frac{21 q_0^5}{16} + \frac{1037 s_0}{2880} - \frac{7 j_0 s_0}{144}
    + \frac{7 q_0 s_0}{12} + \frac{7 q_0^2 s_0}{24} \Bigr) \Bigr]\,,
\end{eqnarray*}
and the angular diameter distance,
\begin{eqnarray*}
    d_A & = & \frac{1}{H_0} \cdot \Bigl[ z + z^2 \cdot \Bigl( -\frac{3}{2} - \frac{q_0}{2} \Bigr)
    + z^3 \cdot \Bigl( \frac{11}{6} -\frac{j_0}{6} + \frac{7 q_0}{6} + \frac{q_0^2}{2} \Bigr) + \nonumber\\
    &&+\, z^4 \cdot \Bigl( -\frac{25}{12} + \frac{13 j_0}{24} - \frac{23 q_0}{12} + \frac{5 j_0 q_0}{12}
    - \frac{13 q_0^2}{8} - \frac{5 q_0^3}{8} + \frac{s_0}{24} \Bigr)  + \nonumber\\
    &&+\, z^5 \cdot \Bigl( \frac{137}{60} - \frac{137 j_0}{120} + \frac{j_0^2}{12} - \frac{l_0}{120}
    + \frac{163 q_0}{60} - \frac{7 j_0 q_0}{4} + \frac{137 q_0^2}{40} - \frac{7 j_0 q_0^2}{8}
    + \frac{21 q_0^3}{8} + \frac{7 q_0^4}{8} - \frac{7 s_0}{40} - \frac{q_0 s_0}{8} \Bigr)  + \nonumber\\
    &&+\, z^6 \cdot \Bigl( -\frac{49}{20} + \frac{79 j_0}{40} - \frac{31 j_0^2}{72} + \frac{31 l_0}{720}
    + \frac{m_0}{720} - \frac{71 q_0}{20} + \frac{163 j_0 q_0}{36} - \frac{7 j_0^2 q_0}{18}
    + \frac{7 l_0 q_0}{240} - \frac{237 q_0^2}{40} + \frac{217 j_0 q_0^2}{48} + \nonumber\\
    &&-\, \frac{163 q_0^3}{24} + \frac{7 j_0 q_0^3}{4} - \frac{217 q_0^4}{48} - \frac{21 q_0^5}{16}
    + \frac{163 s_0}{360} - \frac{7 j_0 s_0}{144} + \frac{31 q_0 s_0}{48} + \frac{7 q_0^2 s_0}{24} \Bigr) \Bigr]\,.
\end{eqnarray*}

\subsection{Redshift $y_1$}
\label{app:y1}
Here we give the results for the distances in terms of $y_1$, first for the luminosity distance,
\begin{eqnarray*}
    d_L & = & \frac{1}{H_0} \cdot \Bigl[ y_1 + y_1^2 \cdot \Bigl(\frac{3}{2} - \frac{q_0}{2} \Bigr)
    + y_1^3 \cdot \Bigl(\frac{11}{6} -\frac{j_0}{6} - \frac{5 q_0}{6} + \frac{q_0^2}{2} \Bigr) + \nonumber\\
    &&+\, y_1^4 \cdot \Bigl( \frac{25}{12} - \frac{7 j_0}{24} - \frac{13 q_0}{12} + \frac{5 j_0 q_0}{12}
    + \frac{7 q_0^2}{8} - \frac{5 q_0^3}{8} + \frac{s_0}{24} \Bigr) + \nonumber\\
    &&+\, y_1^5 \cdot \Bigl( \frac{137}{60} - \frac{47 j_0}{120} + \frac{j_0^2}{12} - \frac{l_0}{120}
    - \frac{77 q_0}{60} + \frac{3 j_0 q_0}{4} + \frac{47 q_0^2}{40} - \frac{7 j_0 q_0^2}{8}
    - \frac{9 q_0^3}{8} + \frac{7 q_0^4}{8} + \frac{3 s_0}{40} - \frac{q_0 s_0}{8} \Bigr) + \nonumber\\
    &&+\, y_1^6 \cdot \Bigl( \frac{49}{20} - \frac{19 j_0}{40} + \frac{11 j_0^2}{72}
    - \frac{11 l_0}{720} + \frac{m_0}{720} - \frac{29 q_0}{20} + \frac{37 j_0 q_0}{36}
    - \frac{7 j_0^2 q_0}{18} + \frac{7 l_0 q_0}{240} + \frac{57 q_0^2}{40} - \frac{77 j_0 q_0^2}{48}
    - \frac{37 q_0^3}{24} + \nonumber\\
    &&+\, \frac{7 j_0 q_0^3}{4} + \frac{77 q_0^4}{48} - \frac{21 q_0^5}{16} + \frac{37 s_0}{360}
    - \frac{7 j_0 s_0}{144} - \frac{11 q_0 s_0}{48} + \frac{7 q_0^2 s_0}{24} \Bigr) \Bigr]\,,
\end{eqnarray*}
the photon flux distance,
\begin{eqnarray*}
    d_F & = & \frac{1}{H_0} \cdot \Bigl[ y_1 + y_1^2 \cdot \Bigl( 1 - \frac{q_0}{2} \Bigr)
    + y_1^3 \cdot \Bigl(\frac{23}{24} -\frac{j_0}{6} - \frac{7 q_0}{12} + \frac{q_0^2}{2} \Bigr) + \nonumber\\
    &&+\, y_1^4 \cdot \Bigl( \frac{11}{12} - \frac{5 j_0}{24} - \frac{29 q_0}{48} + \frac{5 j_0 q_0}{12}
    + \frac{5 q_0^2}{8} - \frac{5 q_0^3}{8} + \frac{s_0}{24} \Bigr)  + \nonumber\\
    &&+\, y_1^5 \cdot \Bigl( \frac{563}{640} - \frac{9 j_0}{40} + \frac{j_0^2}{12} - \frac{l_0}{120}
    - \frac{97 q_0}{160} + \frac{13 j_0 q_0}{24} + \frac{27 q_0^2}{40} - \frac{7 j_0 q_0^2}{8}
    - \frac{13 q_0^3}{16} + \frac{7 q_0^4}{8} + \frac{13 s_0}{240} - \frac{q_0 s_0}{8} \Bigr)  + \nonumber\\
    &&+\, y_1^6 \cdot \Bigl( \frac{1627}{1920} - \frac{223 j_0}{960} + \frac{j_0^2}{9} - \frac{l_0}{90}
    + \frac{m_0}{720} - \frac{2309 q_0}{3840} + \frac{173 j_0 q_0}{288} - \frac{7 j_0^2 q_0}{18}
    + \frac{7 l_0 q_0}{240} + \frac{223 q_0^2}{320} - \frac{7 j_0 q_0^2}{6} - \frac{173 q_0^3}{192} + \nonumber\\
    &&+\, \frac{7 j_0 q_0^3}{4} + \frac{7 q_0^4}{6} - \frac{21 q_0^5}{16} + \frac{173 s_0}{2880}
    - \frac{7 j_0 s_0}{144} - \frac{q_0 s_0}{6} + \frac{7 q_0^2 s_0}{24} \Bigr) \Bigr]\,,
\end{eqnarray*}
the photon count distance,
\begin{eqnarray*}
    d_P & = & \frac{1}{H_0} \cdot \Bigl[ y_1 +y_1^2 \cdot \Bigl( \frac{1}{2} - \frac{q_0}{2} \Bigr)
    + y_1^3 \cdot \Bigl( \frac{1}{3} -\frac{j_0}{6} - \frac{q_0}{3} + \frac{q_0^2}{2} \Bigr) + \nonumber\\
    &&+\, y_1^4 \cdot \Bigl( \frac{1}{4} - \frac{j_0}{8} - \frac{q_0}{4} + \frac{5 j_0 q_0}{12}
    + \frac{3 q_0^2}{8} - \frac{5 q_0^3}{8} + \frac{s_0}{24} \Bigr)  + \nonumber\\
    &&+\, y_1^5 \cdot \Bigl( \frac{1}{5} - \frac{j_0}{10} + \frac{j_0^2}{12} - \frac{l_0}{120}
    - \frac{q_0}{5} + \frac{j_0 q_0}{3} + \frac{3 q_0^2}{10} - \frac{7 j_0 q_0^2}{8} - \frac{q_0^3}{2}
    + \frac{7 q_0^4}{8} + \frac{s_0}{30} - \frac{q_0 s_0}{8} \Bigr)  + \nonumber\\
    &&+\, y_1^6 \cdot \Bigl( \frac{1}{6} - \frac{j_0}{12} + \frac{5 j_0^2}{72} - \frac{l_0}{144}
    + \frac{m_0}{720} - \frac{q_0}{6} + \frac{5 j_0 q_0}{18} - \frac{7 j_0^2 q_0}{18} + \frac{7 l_0 q_0}{240}
    + \frac{q_0^2}{4} - \frac{35 j_0 q_0^2}{48} - \frac{5 q_0^3}{12} + \frac{7 j_0 q_0^3}{4} + \nonumber\\
     &&+\, \frac{35 q_0^4}{48} - \frac{21 q_0^5}{16} + \frac{s_0}{36} - \frac{7 j_0 s_0}{144}
    - \frac{5 q_0 s_0}{48} + \frac{7 q_0^2 s_0}{24} \Bigr) \Bigr]\,,
\end{eqnarray*}
the deceleration distance,
\begin{eqnarray*}
    d_Q & = & \frac{1}{H_0} \cdot \Bigl[ y_1 - y_1^2 \cdot \frac{q_0}{2}
    + y_1^3 \cdot \Bigl( -\frac{1}{24} -\frac{j_0}{6} - \frac{q_0}{12} + \frac{q_0^2}{2} \Bigr) + \nonumber\\
    &&+\, y_1^4 \cdot \Bigl( -\frac{1}{24} - \frac{j_0}{24} - \frac{q_0}{48} + \frac{5 j_0 q_0}{12}
    + \frac{q_0^2}{8} - \frac{5 q_0^3}{8} + \frac{s_0}{24} \Bigr)  + \nonumber\\
    &&+\, y_1^5 \cdot \Bigl( -\frac{71}{1920} - \frac{j_0}{60} + \frac{j_0^2}{12} - \frac{l_0}{120}
    - \frac{q_0}{480} + \frac{j_0 q_0}{8} + \frac{q_0^2}{20} - \frac{7 j_0 q_0^2}{8} - \frac{3 q_0^3}{16}
    + \frac{7 q_0^4}{8} + \frac{s_0}{80} - \frac{q_0 s_0}{8} \Bigr)  + \nonumber\\
    &&+\, y_1^6 \cdot \Bigl( -\frac{31}{960} - \frac{7 j_0}{960} + \frac{j_0^2}{36} - \frac{l_0}{360}
    + \frac{m_0}{720} + \frac{19 q_0}{3840} + \frac{17 j_0 q_0}{288} - \frac{7 j_0^2 q_0}{18}
    + \frac{7 l_0 q_0}{240} + \frac{7 q_0^2}{320} - \frac{7 j_0 q_0^2}{24} - \frac{17 q_0^3}{192} + \nonumber\\
    &&+\, \frac{7 j_0 q_0^3}{4} + \frac{7 q_0^4}{24} - \frac{21 q_0^5}{16} + \frac{17 s_0}{2880}
    - \frac{7 j_0 s_0}{144} - \frac{q_0 s_0}{24} + \frac{7 q_0^2 s_0}{24} \Bigr) \Bigr]\,,
\end{eqnarray*}
and the angular diameter distance,
\begin{eqnarray*}
    d_A & = & \frac{1}{H_0} \cdot \Bigl[ y_1 + y_1^2 \cdot \Bigl( -\frac{1}{2} - \frac{q_0}{2} \Bigr)
    + y_1^3 \cdot \Bigl( -\frac{1}{6} -\frac{j_0}{6} + \frac{q_0}{6} + \frac{q_0^2}{2} \Bigr) + \nonumber\\
    &&+\, y_1^4 \cdot \Bigl( -\frac{1}{12} + \frac{j_0}{24} + \frac{q_0}{12} + \frac{5 j_0 q_0}{12}
    - \frac{q_0^2}{8} - \frac{5 q_0^3}{8} + \frac{s_0}{24} \Bigr)  + \nonumber\\
    &&+\, y_1^5 \cdot \Bigl( -\frac{1}{20} + \frac{j_0}{40} + \frac{j_0^2}{12} - \frac{l_0}{120}
    + \frac{q_0}{20} - \frac{j_0 q_0}{12} - \frac{3 q_0^2}{40} - \frac{7 j_0 q_0^2}{8} + \frac{q_0^3}{8}
    + \frac{7 q_0^4}{8} - \frac{s_0}{120} - \frac{q_0 s_0}{8} \Bigr)  + \nonumber\\
    &&+\, y_1^6 \cdot \Bigl( -\frac{1}{30} + \frac{j_0}{60} - \frac{j_0^2}{72} + \frac{l_0}{720}
    + \frac{m_0}{720} + \frac{q_0}{30} - \frac{j_0 q_0}{18} - \frac{7 j_0^2 q_0}{18} + \frac{7 l_0 q_0}{240}
    - \frac{q_0^2}{20} + \frac{7 j_0 q_0^2}{48} + \frac{q_0^3}{12} + \frac{7 j_0 q_0^3}{4} + \nonumber\\
    &&-\, \frac{7 q_0^4}{48} - \frac{21 q_0^5}{16} - \frac{s_0}{180} - \frac{7 j_0 s_0}{144}
    + \frac{q_0 s_0}{48} + \frac{7 q_0^2 s_0}{24} \Bigr) \Bigr]\,.
\end{eqnarray*}

\subsection{Redshift $y_2$}
\label{app:y2}
The cosmological distances are given in terms of $y_2$ - luminosity distance,
\begin{eqnarray*}
    d_L & = & \frac{1}{H_0} \cdot \Bigl[ y_2 + y_2^2 \cdot \Bigl(\frac{3}{2} - \frac{q_0}{2} \Bigr)
    + y_2^3 \cdot \Bigl(\frac{13}{6} -\frac{j_0}{6} - \frac{5q_0}{6} + \frac{q_0^2}{2} \Bigr) + \nonumber\\
    &&+\, y_2^4 \cdot \Bigl( \frac{37}{12} - \frac{7 j_0}{24} - \frac{17 q_0}{12} + \frac{5 j_0 q_0}{12}
    - \frac{7 q_0^2}{8} - \frac{5 q_0^3}{8} + \frac{s}{24} \Bigr) + \nonumber\\
    &&+\, y_2^5 \cdot \Bigl( \frac{17}{4} - \frac{67 j_0}{120} + \frac{j_0^2}{12} - \frac{l_0}{120}
    - \frac{127 q_0}{60} + \frac{3 j_0 q_0}{4} + \frac{67 q_0^2}{40} - \frac{7 j_0 q_0^2}{8}
    - \frac{9 q_0^3}{8} + \frac{7 q_0^4}{8} + \frac{3 s_0}{40} - \frac{q_0 s_0}{8} \Bigr) + \nonumber\\
    &&+\, y_2^6 \cdot \Bigl( \frac{1043}{180} - \frac{311 j_0}{360} + \frac{11 j_0^2}{72}
    - \frac{11 l_0}{720} + \frac{m_0}{720} - \frac{37 q_0}{12} + \frac{37 j_0 q_0}{9}
    - \frac{7 j_0^2 q_0}{36} - \frac{7 l_0 q_0}{240}- \frac{89 q_0^2}{120} + \frac{21 j_0 q_0^2}{16} + \nonumber\\
    &&+\, \frac{19 q_0^3}{8} + \frac{7 j_0 q_0^3}{4} + \frac{77 q_0^4}{48} - \frac{21 q_0^5}{16}
    + \frac{19 s_0}{120} - \frac{7 j_0 s_0}{144} - \frac{11 q_0 s_0}{48} + \frac{7 q_0^2 s_0}{24} \Bigr) \Bigr]\,,
\end{eqnarray*}
the photon flux distance,
\begin{eqnarray*}
    d_F & = & \frac{1}{H_0} \cdot \Bigl[ y_2 +y_2^2 \cdot \Bigl( 1 - \frac{q_0}{2} \Bigr)
    + y_2^3 \cdot \Bigl(\frac{31}{24} -\frac{j_0}{6} - \frac{7q_0}{12} + \frac{q_0^2}{2} \Bigr) + \nonumber\\
    &&+\, y_2^4 \cdot \Bigl( \frac{19}{12} - \frac{5 j_0}{24} - \frac{15 q_0}{16} + \frac{5 j_0 q_0}{12}
    + \frac{5 q_0^2}{8} - \frac{5 q_0^3}{8} + \frac{s_0}{24} \Bigr)  + \nonumber\\
    &&+\, y_2^5 \cdot \Bigl( \frac{757}{384} - \frac{47 j_0}{120} + \frac{j_0^2}{12}
    - \frac{l_0}{120} - \frac{571 q_0}{480} + \frac{13 j_0 q_0}{24} + \frac{47 q_0^2}{40}
    - \frac{7 j_0 q_0^2}{8} - \frac{13 q_0^3}{16} + \frac{7 q_0^4}{8} + \frac{13 s_0}{240}
    - \frac{q_0 s_0}{8} \Bigr)  + \nonumber\\
    &&+\, y_2^6 \cdot \Bigl( \frac{4699}{1920} - \frac{1469 j_0}{2880} + \frac{j_0^2}{9}
    - \frac{l_0}{90} + \frac{m_0}{720} - \frac{18383 q_0}{11520} + \frac{37 j_0 q_0}{32}
    - \frac{7 j_0^2 q_0}{18} + \frac{7 l_0 q_0}{240} + \frac{1469 q_0^2}{960} - \frac{7 j_0 q_0^2}{6} + \nonumber\\
    &&-\, \frac{111 q_0^3}{64} + \frac{7 j_0 q_0^3}{4} + \frac{7 q_0^4}{6} - \frac{21 q_0^5}{16}
    + \frac{37 s_0}{320} - \frac{7 j_0 s_0}{144} - \frac{q_0 s_0}{6} + \frac{7 q_0^2 s_0}{24} \Bigr) \Bigr]\,,
\end{eqnarray*}
the photon count distance,
\begin{eqnarray*}
    d_P & = & \frac{1}{H_0} \cdot \Bigl[ y_2 +y_2^2 \cdot \Bigl( \frac{1}{2} - \frac{q_0}{2} \Bigr)
    + y_2^3 \cdot \Bigl( \frac{2}{3} -\frac{j_0}{6} - \frac{q_0}{3} + \frac{q_0^2}{2} \Bigr) + \nonumber\\
    &&+\, y_2^4 \cdot \Bigl( \frac{7}{12} - \frac{j_0}{8} - \frac{7 q_0}{12} + \frac{5 j_0 q_0}{12}
    + \frac{3 q_0^2}{8} - \frac{5 q_0^3}{8} + \frac{s_0}{24} \Bigr)  + \nonumber\\
    &&+\, y_2^5 \cdot \Bigl( \frac{2}{3} - \frac{4 j_0}{15} + \frac{j_0^2}{12} - \frac{l_0}{120}
    - \frac{8 q_0}{15} + \frac{j_0 q_0}{3} + \frac{4 q_0^2}{5} - \frac{7 j_0 q_0^2}{8} - \frac{q_0^3}{2}
    + \frac{7 q_0^4}{8} + \frac{s_0}{30} - \frac{q_0 s_0}{8} \Bigr)  + \nonumber\\
    &&+\, y_2^6 \cdot \Bigl( \frac{31}{45} - \frac{j_0}{4} + \frac{5 j_0^2}{72} - \frac{l_0}{144}
    + \frac{m_0}{720} - \frac{31 q_0}{45} + \frac{5 j_0 q_0}{6} - \frac{7 j_0^2 q_0}{18}
    + \frac{7 l_0 q_0}{240} + \frac{3 q_0^2}{4} - \frac{35 j_0 q_0^2}{48} - \frac{5 q_0^3}{4} + \nonumber\\
    &&+\, \frac{7 j_0 q_0^3}{4} + \frac{35 q_0^4}{48} - \frac{21 q_0^5}{16} + \frac{s_0}{12}
    - \frac{7 j_0 s_0}{144} - \frac{5 q_0 s_0}{48} + \frac{7 q_0^2 s_0}{24} \Bigr) \Bigr]\,,
\end{eqnarray*}
the deceleration distance,
\begin{eqnarray*}
    d_Q & = & \frac{1}{H_0} \cdot \Bigl[ y_2 - y_2^2 \cdot \frac{q_0}{2}
    + y_2^3 \cdot \Bigl( \frac{7}{24} -\frac{j_0}{6} - \frac{q_0}{12} + \frac{q_0^2}{2} \Bigr) + \nonumber\\
    &&+\, y_2^4 \cdot \Bigl( -\frac{1}{24} - \frac{j_0}{24} - \frac{17 q_0}{48} + \frac{5 j_0 q_0}{12}
    + \frac{q_0^2}{8} - \frac{5 q_0^3}{8} + \frac{s_0}{24} \Bigr)  + \nonumber\\
    &&+\, y_2^5 \cdot \Bigl( \frac{7}{128} - \frac{11 j_0}{60} + \frac{j_0^2}{12} - \frac{l_0}{120}
    - \frac{41 q_0}{480} + \frac{j_0 q_0}{8} + \frac{11 q_0^2}{20} - \frac{7 j_0 q_0^2}{8}
    - \frac{3 q_0^3}{16} + \frac{7 q_0^4}{8} + \frac{s_0}{80} - \frac{q_0 s_0}{8} \Bigr)  + \nonumber\\
    &&+\, y_2^6 \cdot \Bigl( -\frac{253}{2880} - \frac{181 j_0}{2880} + \frac{j_0^2}{36} - \frac{l_0}{360}
    + \frac{m_0}{720} - \frac{271 q_0}{1280} + \frac{59 j_0 q_0}{96} - \frac{7 j_0^2 q_0}{18}
    + \frac{7 l_0 q_0}{240} + \frac{181 q_0^2}{960} - \frac{7 j_0 q_0^2}{24} + \nonumber\\
    &&-\, \frac{59 q_0^3}{64} + \frac{7 j_0 q_0^3}{4} + \frac{7 q_0^4}{24} - \frac{21 q_0^5}{16}
    + \frac{59 s_0}{960} - \frac{7 j_0 s_0}{144} - \frac{q_0 s_0}{24} + \frac{7 q_0^2 s_0}{24} \Bigr) \Bigr]\,,
\end{eqnarray*}
and the angular diameter distance,
\begin{eqnarray*}
    d_A & = & \frac{1}{H_0} \cdot \Bigl[ y_2 + y_2^2 \cdot \Bigl( -\frac{1}{2} - \frac{q_0}{2} \Bigr)
    + y_2^3 \cdot \Bigl( \frac{1}{6} -\frac{j_0}{6} + \frac{q_0}{6} + \frac{q_0^2}{2} \Bigr) + \nonumber\\
    &&+\, y_2^4 \cdot \Bigl( -\frac{5}{12} + \frac{j_0}{24} - \frac{q_0}{4} + \frac{5 j_0 q_0}{12}
    - \frac{q_0^2}{8} - \frac{5 q_0^3}{8} + \frac{s_0}{24} \Bigr)  + \nonumber\\
    &&+\, y_2^5 \cdot \Bigl( -\frac{1}{12} - \frac{17 j_0}{120} + \frac{j_0^2}{12} - \frac{l_0}{120}
    + \frac{13 q_0}{60} - \frac{j_0 q_0}{12} + \frac{17 q_0^2}{40} - \frac{7 j_0 q_0^2}{8} + \frac{q_0^3}{8}
    + \frac{7 q_0^4}{8} - \frac{s_0}{120} - \frac{q_0 s_0}{8} \Bigr)  + \nonumber\\
    &&+\, y_2^6 \cdot \Bigl( -\frac{1}{3} + \frac{13 j_0}{180} - \frac{j_0^2}{72} + \frac{l_0}{720}
    + \frac{m_0}{720} - \frac{2 q_0}{45} + \frac{j_0 q_0}{2} - \frac{7 j_0^2 q_0}{18} + \frac{7 l_0 q_0}{240}
    - \frac{13 q_0^2}{60} + \frac{7 j_0 q_0^2}{48} - \frac{3 q_0^3}{4} + \nonumber\\
    &&+\, \frac{7 j_0 q_0^3}{4} - \frac{7 q_0^4}{48} - \frac{21 q_0^5}{16} + \frac{s_0}{20}
    - \frac{7 j_0 s_0}{144} + \frac{q_0 s_0}{48} + \frac{7 q_0^2 s_0}{24} \Bigr) \Bigr]\,.
\end{eqnarray*}

\subsection{Redshift $y_3$}
\label{app:y3}
These are the results for the cosmological distances in terms of the third redshift, $y_3$,
beginning with the luminosity distance,
\begin{eqnarray*}
    d_L & = & \frac{1}{H_0} \cdot \Bigl[ y_3 + y_3^2 \cdot \Bigl(\frac{1}{2} - \frac{q_0}{2} \Bigr)
    + y_3^3 \cdot \Bigl(\frac{5}{6} -\frac{j_0}{6} + \frac{q_0}{6} + \frac{q_0^2}{2} \Bigr) + \nonumber\\
    &&+\, y_3^4 \cdot \Bigl( \frac{13}{12} + \frac{5 j_0}{24} - \frac{13 q_0}{12} + \frac{5 j_0 q_0}{12}
    - \frac{5 q_0^2}{8} - \frac{5 q_0^3}{8} + \frac{s}{24} \Bigr) + \nonumber\\
    &&+\, y_3^5 \cdot \Bigl( \frac{29}{20} - \frac{29 j_0}{40} + \frac{j_0^2}{12} - \frac{l_0}{120}
    + \frac{11 q_0}{20} - \frac{11 j_0 q_0}{12} - \frac{87 q_0^2}{40} - \frac{7 j_0 q_0^2}{8}
    + \frac{11 q_0^3}{8} + \frac{7 q_0^4}{8} - \frac{11 s_0}{120} - \frac{q_0 s_0}{8} \Bigr) + \nonumber\\
    &&+\, y_3^6 \cdot \Bigl( \frac{43}{15} + \frac{16 j_0}{15} - \frac{19 j_0^2}{72} + \frac{19 l_0}{720}
    + \frac{m_0}{720} - \frac{43 q_0}{15} + \frac{28 j_0 q_0}{9} - \frac{7 j_0^2 q_0}{18}
    + \frac{7 l_0 q_0}{240} - \frac{16 q_0^2}{5} + \frac{133 j_0 q_0^2}{48} - \frac{14 q_0^3}{3} + \nonumber\\
    &&+\, \frac{7 j_0 q_0^3}{4} - \frac{133 q_0^4}{48} - \frac{21 q_0^5}{16}
    + \frac{14 s_0}{45} - \frac{7 j_0 s_0}{144} + \frac{19 q_0 s_0}{48} + \frac{7 q_0^2 s_0}{24} \Bigr) \Bigr]\,,
\end{eqnarray*}
the photon flux distance,
\begin{eqnarray*}
    d_F & = & \frac{1}{H_0} \cdot \Bigl[ y_3 - y_3^2 \cdot \frac{q_0}{2}
    + y_3^3 \cdot \Bigl(\frac{23}{24} -\frac{j_0}{6} + \frac{5q_0}{12} + \frac{q_0^2}{2} \Bigr) + \nonumber\\
    &&+\, y_3^4 \cdot \Bigl( \frac{1}{24} + \frac{7 j_0}{24} - \frac{65 q_0}{48} + \frac{5 j_0 q_0}{12}
    - \frac{7 q_0^2}{8} - \frac{5 q_0^3}{8} + \frac{s_0}{24} \Bigr)  + \nonumber\\
    &&+\, y_3^5 \cdot \Bigl( \frac{3529}{1920} - \frac{107 j_0}{120} + \frac{j_0^2}{12}
    - \frac{l_0}{120} + \frac{749 q_0}{480} - \frac{9 j_0 q_0}{8} + \frac{107 q_0^2}{40}
    - \frac{7 j_0 q_0^2}{8} + \frac{27 q_0^3}{16} + \frac{7 q_0^4}{8} - \frac{9 s_0}{80}
    - \frac{q_0 s_0}{8} \Bigr)  + \nonumber\\
    &&+\, y_3^6 \cdot \Bigl( \frac{191}{960} + \frac{1577 j_0}{960} - \frac{11 j_0^2}{36}
    + \frac{11 l_0}{360} + \frac{m_0}{720} - \frac{16109 q_0}{3840} + \frac{1073 j_0 q_0}{288}
    - \frac{7 j_0^2 q_0}{18} + \frac{7 l_0 q_0}{240} - \frac{1577 q_0^2}{320} + \nonumber\\
    &&+\, \frac{77 j_0 q_0^2}{24}
    - \frac{1073 q_0^3}{192} + \frac{7 j_0 q_0^3}{4} - \frac{77 q_0^4}{24} - \frac{21 q_0^5}{16}
    + \frac{1073 s_0}{2880} - \frac{7 j_0 s_0}{144} + \frac{11 q_0 s_0}{24} + \frac{7 q_0^2 s_0}{24} \Bigr) \Bigr]\,,
\end{eqnarray*}
the photon count distance,
\begin{eqnarray*}
    d_P & = & \frac{1}{H_0} \cdot \Bigl[ y_3 +y_3^2 \cdot \Bigl( -\frac{1}{2} - \frac{q_0}{2} \Bigr)
    + y_3^3 \cdot \Bigl( \frac{4}{3} -\frac{j_0}{6} + \frac{2q_0}{3} + \frac{q_0^2}{2} \Bigr) + \nonumber\\
    &&+\, y_3^4 \cdot \Bigl( -\frac{5}{4} + \frac{3 j_0}{8} - \frac{7 q_0}{4} + \frac{5 j_0 q_0}{12}
    - \frac{9 q_0^2}{8} - \frac{5 q_0^3}{8} + \frac{s_0}{24} \Bigr) + \nonumber\\
    &&+\, y_3^5 \cdot \Bigl( \frac{16}{5} - \frac{11 j_0}{10} + \frac{j_0^2}{12} - \frac{l_0}{120}
    + \frac{14 q_0}{5} - \frac{4 j_0 q_0}{3} + \frac{33 q_0^2}{10} - \frac{7 j_0 q_0^2}{8} + 2 q_0^3
    + \frac{7 q_0^4}{8} - \frac{2 s_0}{15} - \frac{q_0 s_0}{8} \Bigr)  + \nonumber\\
    &&+\, y_3^6 \cdot \Bigl( -\frac{11}{3} + \frac{7 j_0}{3} - \frac{25 j_0^2}{72} + \frac{5 l_0}{144}
    + \frac{m_0}{720} - \frac{19 q_0}{3} + \frac{40 j_0 q_0}{9} - \frac{7 j_0^2 q_0}{18} +
    + \frac{7 l_0 q_0}{240} - 7 q_0^2 + \nonumber\\
    &&+\, \frac{175 j_0 q_0^2}{48} - \frac{20 q_0^3}{3}
    + \frac{7 j_0 q_0^3}{4} - \frac{175 q_0^4}{48} - \frac{21 q_0^5}{16} + \frac{4s_0}{9}
    - \frac{7 j_0 s_0}{144} + \frac{25 q_0 s_0}{48} + \frac{7 q_0^2 s_0}{24} \Bigr) \Bigr]\,,
\end{eqnarray*}
the deceleration distance,
\begin{eqnarray*}
    d_Q & = & \frac{1}{H_0} \cdot \Bigl[ y_3 +y_3^2 \cdot \Bigl( -1 - \frac{q_0}{2} \Bigr)
    + y_3^3 \cdot \Bigl( \frac{47}{24} -\frac{j_0}{6} + \frac{11 q_0}{12} + \frac{q_0^2}{2} \Bigr) + \nonumber\\
    &&+\, y_3^4 \cdot \Bigl( -\frac{35}{12} + \frac{11 j_0}{24} - \frac{109 q_0}{48} + \frac{5 j_0 q_0}{12}
    - \frac{11 q_0^2}{8} - \frac{5 q_0^3}{8} + \frac{s_0}{24} \Bigr)  + \nonumber\\
    &&+\, y_3^5 \cdot \Bigl( \frac{3683}{640} - \frac{27 j_0}{20} + \frac{j_0^2}{12} - \frac{l_0}{120}
    + \frac{693 q_0}{160} - \frac{37 j_0 q_0}{24} + \frac{81 q_0^2}{20} - \frac{7 j_0 q_0^2}{8}
    + \frac{37 q_0^3}{16} + \frac{7 q_0^4}{8} - \frac{37 s_0}{240} - \frac{q_0 s_0}{8} \Bigr)  + \nonumber\\
    &&+\, y_3^6 \cdot \Bigl( -\frac{6089}{640} + \frac{1011 j_0}{320} - \frac{7 j_0^2}{18}
    + \frac{7 l_0}{18} + \frac{m_0}{720} - \frac{12087 q_0}{1280} + \frac{1517 j_0 q_0}{288}
    - \frac{7 j_0^2 q_0}{18} + \frac{7 l_0 q_0}{240} - \frac{3033 q_0^2}{320} + \nonumber\\
    &&+\, \frac{49 j_0 q_0^2}{12} - \frac{1517 q_0^3}{192}
    + \frac{7 j_0 q_0^3}{4} - \frac{49 q_0^4}{12} - \frac{21 q_0^5}{16} + \frac{1517 s_0}{2880}
    - \frac{7 j_0 s_0}{144} + \frac{7 q_0 s_0}{12} + \frac{7 q_0^2 s_0}{24} \Bigr) \Bigr]\,,
\end{eqnarray*}
and the angular diameter distance,
\begin{eqnarray*}
    d_A & = & \frac{1}{H_0} \cdot \Bigl[ y_3 +y_3^2 \cdot \Bigl( -\frac{3}{2} - \frac{q_0}{2} \Bigr)
    + y_3^3 \cdot \Bigl( \frac{17}{6} -\frac{j_0}{6} + \frac{7 q_0}{6} + \frac{q_0^2}{2} \Bigr) + \nonumber\\
    &&+\, y_3^4 \cdot \Bigl( -\frac{61}{12} + \frac{13 j_0}{24} - \frac{35 q_0}{12} + \frac{5 j_0 q_0}{12}
    - \frac{13 q_0^2}{8} - \frac{5 q_0^3}{8} + \frac{s_0}{24} \Bigr)  + \nonumber\\
    &&+\, y_3^5 \cdot \Bigl( \frac{587}{60} - \frac{197 j_0}{120} + \frac{j_0^2}{12} - \frac{l_0}{120}
    + \frac{373 q_0}{60} - \frac{7 j_0 q_0}{4} + \frac{197 q_0^2}{40} - \frac{7 j_0 q_0^2}{8}
    + \frac{21 q_0^3}{8} + \frac{7 q_0^4}{8} - \frac{7 s_0}{40} - \frac{q_0 s_0}{8} \Bigr)  + \nonumber\\
    &&+\, y_3^6 \cdot \Bigl( -\frac{1097}{60} + \frac{497 j_0}{120} - \frac{31 j_0^2}{72}
    + \frac{31 l_0}{720} + \frac{m_0}{720} - \frac{823 q_0}{60} + \frac{223 j_0 q_0}{36}
    - \frac{7 j_0^2 q_0}{18} + \frac{7 l_0 q_0}{240} - \frac{497 q_0^2}{40} + \nonumber\\
    &&+\, \frac{217 j_0 q_0^2}{48}
    - \frac{223 q_0^3}{24} + \frac{7 j_0 q_0^3}{4} - \frac{217 q_0^4}{48} - \frac{21 q_0^5}{16}
    + \frac{223 s_0}{360} - \frac{7 j_0 s_0}{144} + \frac{31 q_0 s_0}{48} + \frac{7 q_0^2 s_0}{24} \Bigr) \Bigr]\,.
\end{eqnarray*}

\subsection{Redshift $y_4$}
\label{app:y4}
Finally, for the last redshift $y_4$ the cosmological distances are given, starting with the
luminosity distance,
\begin{eqnarray*}
    d_L & = & \frac{1}{H_0} \cdot \Bigl[ y_4 + y_4^2 \cdot \Bigl(\frac{1}{2} - \frac{q_0}{2} \Bigr)
    + y_4^3 \cdot \Bigl(\frac{1}{6} -\frac{j_0}{6} + \frac{q_0}{6} + \frac{q_0^2}{2} \Bigr) + \nonumber\\
    &&+\, y_4^4 \cdot \Bigl( \frac{5}{12} + \frac{5 j_0}{24} - \frac{5 q_0}{12} + \frac{5 j_0 q_0}{12}
    - \frac{5 q_0^2}{8} - \frac{5 q_0^3}{8} + \frac{s}{24} \Bigr) + \nonumber\\
    &&+\, y_4^5 \cdot \Bigl( -\frac{1}{12} - \frac{47 j_0}{120} + \frac{j_0^2}{12} - \frac{l_0}{120}
    + \frac{13 q_0}{60} - \frac{11 j_0 q_0}{12} + \frac{47 q_0^2}{40} - \frac{7 j_0 q_0^2}{8} + \frac{11 q_0^3}{8}
    + \frac{7 q_0^4}{8} - \frac{11 s_0}{120} - \frac{q_0 s_0}{8} \Bigr) + \nonumber\\
    &&+\, y_4^6 \cdot \Bigl( \frac{1}{3} + \frac{23 j_0}{45} - \frac{19 j_0^2}{72} + \frac{19 l_0}{720}
    + \frac{m_0}{720} - \frac{q_0}{3} + 2 j_0 q_0 - \frac{7 j_0^2 q_0}{18} + \frac{7 l_0 q_0}{240}
    - \frac{23 q_0^2}{15} + \frac{133 j_0 q_0^2}{48} - 3 q_0^3 + \nonumber\\
    &&+\, \frac{7 j_0 q_0^3}{4} - \frac{133 q_0^4}{48} - \frac{21 q_0^5}{16} + \frac{s_0}{5}
    - \frac{7 j_0 s_0}{144} + \frac{19 q_0 s_0}{48} + \frac{7 q_0^2 s_0}{24} \Bigr) \Bigr]\,,
\end{eqnarray*}
the photon flux distance,
\begin{eqnarray*}
    d_F & = & \frac{1}{H_0} \cdot \Bigl[ y_4 - y_4^2 \cdot \frac{q_0}{2}
    + y_4^3 \cdot \Bigl(\frac{7}{24} -\frac{j_0}{6} + \frac{5q_0}{12} + \frac{q_0^2}{2} \Bigr) + \nonumber\\
    &&+\, y_4^4 \cdot \Bigl( \frac{1}{24} + \frac{7 j_0}{24} - \frac{11 q_0}{16} + \frac{5 j_0 q_0}{12}
    - \frac{7 q_0^2}{8} - \frac{5 q_0^3}{8} + \frac{s_0}{24} \Bigr)  + \nonumber\\
    &&+\, y_4^5 \cdot \Bigl( \frac{7}{128} - \frac{67 j_0}{120} + \frac{j_0^2}{12}
    - \frac{l_0}{120} + \frac{349 q_0}{480} - \frac{9 j_0 q_0}{8} + \frac{67 q_0^2}{40}
    - \frac{7 j_0 q_0^2}{8} + \frac{27 q_0^3}{16} + \frac{7 q_0^4}{8} - \frac{9 s_0}{80}
    - \frac{q_0 s_0}{8} \Bigr)  + \nonumber\\
    &&+\, y_4^6 \cdot \Bigl( \frac{253}{2880} + \frac{2491 j_0}{2880} - \frac{11 j_0^2}{36}
    + \frac{11 l_0}{360} + \frac{m_0}{720} - \frac{10823 q_0}{11520} + \frac{251 j_0 q_0}{96}
    - \frac{7 j_0^2 q_0}{18} + \frac{7 l_0 q_0}{240} - \frac{2491 q_0^2}{960} + \nonumber\\
    &&+\, \frac{77 j_0 q_0^2}{24} - \frac{251 q_0^3}{64} + \frac{7 j_0 q_0^3}{4} - \frac{77 q_0^4}{24}
    - \frac{21 q_0^5}{16} + \frac{251 s_0}{960} - \frac{7 j_0 s_0}{144} + \frac{11 q_0 s_0}{24}
    + \frac{7 q_0^2 s_0}{24} \Bigr) \Bigr]\,,
\end{eqnarray*}
the photon count distance,
\begin{eqnarray*}
    d_P & = & \frac{1}{H_0} \cdot \Bigl[ y_4 +y_4^2 \cdot \Bigl( -\frac{1}{2} - \frac{q_0}{2} \Bigr) +
    y_4^3 \cdot \Bigl( \frac{2}{3} -\frac{j_0}{6} + \frac{2q_0}{3} + \frac{q_0^2}{2} \Bigr) + \nonumber\\
    &&+\, y_4^4 \cdot \Bigl( -\frac{7}{12} + \frac{3 j_0}{8} - \frac{13 q_0}{12} + \frac{5 j_0 q_0}{12} -
    \frac{9 q_0^2}{8} - \frac{5 q_0^3}{8} + \frac{s_0}{24} \Bigr) + \nonumber\\
    &&+\, y_4^5 \cdot \Bigl( \frac{2}{3} - \frac{23 j_0}{30} + \frac{j_0^2}{12} - \frac{l_0}{120} + \frac{22 q_0}{15}
    - \frac{4 j_0 q_0}{3} + \frac{23 q_0^2}{10} - \frac{7 j_0 q_0^2}{8} + 2 q_0^3 + \frac{7 q_0^4}{8}
    - \frac{2 s_0}{15} - \frac{q_0 s_0}{8} \Bigr)  + \nonumber\\
    &&+\, y_4^6 \cdot \Bigl( -\frac{31}{45} + \frac{4 j_0}{3} - \frac{25 j_0^2}{72} + \frac{5 l_0}{144} + \frac{m_0}{720}
    - \frac{91 q_0}{45} + \frac{10 j_0 q_0}{3} - \frac{7 j_0^2 q_0}{18} + \frac{7 l_0 q_0}{240} - 4 q_0^2
    + \frac{175 j_0 q_0^2}{48} - 5 q_0^3 + \nonumber\\
    &&+\, \frac{7 j_0 q_0^3}{4} - \frac{175 q_0^4}{48} - \frac{21 q_0^5}{16}
    + \frac{s_0}{3} - \frac{7 j_0 s_0}{144} + \frac{25 q_0 s_0}{48} + \frac{7 q_0^2 s_0}{24} \Bigr) \Bigr]\,,
\end{eqnarray*}
the deceleration distance,
\begin{eqnarray*}
    d_Q & = & \frac{1}{H_0} \cdot \Bigl[ y_4 +y_4^2 \cdot \Bigl( -1 - \frac{q_0}{2} \Bigr) +
    y_4^3 \cdot \Bigl( \frac{31}{24} -\frac{j_0}{6} + \frac{11 q_0}{12} + \frac{q_0^2}{2} \Bigr) + \nonumber\\
    &&+\, y_4^4 \cdot \Bigl( -\frac{19}{12} + \frac{11 j_0}{24} - \frac{77 q_0}{48} + \frac{5 j_0 q_0}{12}
    - \frac{11 q_0^2}{8} - \frac{5 q_0^3}{8} + \frac{s_0}{24} \Bigr)  + \nonumber\\
    &&+\, y_4^5 \cdot \Bigl( \frac{757}{384} - \frac{61 j_0}{60} + \frac{j_0^2}{12} - \frac{l_0}{120}
    + \frac{1199 q_0}{480} - \frac{37 j_0 q_0}{24} + \frac{61 q_0^2}{20} - \frac{7 j_0 q_0^2}{8} + \frac{37 q_0^3}{16} +
    \frac{7 q_0^4}{8} - \frac{37 s_0}{240} - \frac{q_0 s_0}{8} \Bigr)  + \nonumber\\
    &&+\, y_4^6 \cdot \Bigl( -\frac{4699}{1920} + \frac{5579 j_0}{2880} - \frac{7 j_0^2}{18} + \frac{7 l_0}{180}
    + \frac{m_0}{720} - \frac{4791 q_0}{1280} + \frac{133 j_0 q_0}{32} - \frac{7 j_0^2 q_0}{18} + \frac{7 l_0 q_0}{240}
    - \frac{5579 q_0^2}{960} + \nonumber\\
    &&+\, \frac{49 j_0 q_0^2}{12} - \frac{399 q_0^3}{64} + \frac{7 j_0 q_0^3}{4} - \frac{49 q_0^4}{12}
    - \frac{21 q_0^5}{16} + \frac{133 s_0}{320} - \frac{7 j_0 s_0}{144} + \frac{7 q_0 s_0}{12}
    + \frac{7 q_0^2 s_0}{24} \Bigr) \Bigr]\,,
\end{eqnarray*}
and the angular diameter distance,
\begin{eqnarray*}
    d_A & = & \frac{1}{H_0} \cdot \Bigl[ y_4 +y_4^2 \cdot \Bigl( -\frac{3}{2} - \frac{q_0}{2} \Bigr) +
    y_4^3 \cdot \Bigl( \frac{13}{6} -\frac{j_0}{6} + \frac{7 q_0}{6} + \frac{q_0^2}{2} \Bigr) + \nonumber\\
    &&+\, y_4^4 \cdot \Bigl( -\frac{37}{12} + \frac{13 j_0}{24} - \frac{9 q_0}{4} + \frac{5 j_0 q_0}{12}
    - \frac{13 q_0^2}{8} - \frac{5 q_0^3}{8} + \frac{s_0}{24} \Bigr)  + \nonumber\\
    &&+\, y_4^5 \cdot \Bigl( \frac{17}{4} - \frac{157 j_0}{120} + \frac{j_0^2}{12} - \frac{l_0}{120}
    + \frac{233 q_0}{60} - \frac{7 j_0 q_0}{4} + \frac{157 q_0^2}{40} - \frac{7 j_0 q_0^2}{8}
    + \frac{21 q_0^3}{8} + \frac{7 q_0^4}{8} - \frac{7 s_0}{40} - \frac{q_0 s_0}{8} \Bigr)  + \nonumber\\
    &&+\, y_4^6 \cdot \Bigl( -\frac{1043}{180} + \frac{971 j_0}{360} - \frac{31 j_0^2}{72}
    + \frac{31 l_0}{720} + \frac{m_0}{720} - \frac{1133 q_0}{180} + \frac{61 j_0 q_0}{12} - \frac{7 j_0^2 q_0}{18} +
    + \frac{7 l_0 q_0}{240} - \frac{971 q_0^2}{120} + \nonumber\\
    &&+\, \frac{217 j_0 q_0^2}{48} - \frac{61 q_0^3}{8}
    + \frac{7 j_0 q_0^3}{4} - \frac{217 q_0^4}{48} - \frac{21 q_0^5}{16} + \frac{61 s_0}{120}
    - \frac{7 j_0 s_0}{144} + \frac{31 q_0 s_0}{48} + \frac{7 q_0^2 s_0}{24} \Bigr) \Bigr]\,.
\end{eqnarray*}

\section{The Hubble parameter as a function of redshifts}
\label{app:Hubble}
We start with the parametrization of the Hubble parameter in terms of
the commonly used redshift $z$, given by
\begin{eqnarray*}
    H(z) & = & H_0 \cdot \Bigl[ 1 + z\cdot \Bigl( 1+q_0 \Bigr) + \frac{z^2}{2} \cdot \Bigl( j_0 - q_0^2 \Bigr) +
    \frac{z^3}{6} \cdot \Bigl( -3 q_0^2 - 3 q_0^3 + j_0 \bigl(3 + 4 q_0\bigr) + s_0 \Bigr) + \nonumber\\
    &&+\, \frac{z^4}{24} \cdot \Bigl( -4 j_0^2 + l_0 - 12 q_0^2 - 24 q_0^3 - 15 q_0^4 + j_0 \bigl(12 + 32 q_0 +
    25 q_0^2\bigr) + 8 s_0 + 7 q_0 s_0 \Bigr) + \nonumber\\
    &&+\, \frac{z^5}{120} \cdot \Bigl( m_0 - 60 q_0^2 - 180 q_0^3 - 225 q_0^4 - 105 q_0^5 - 10 j_0^2
    \bigl(6 + 7 q_0\bigr) + l_0 \bigl(15 + 11 q_0\bigr) + \nonumber\\
    &&+\, 15 j_0 \bigl(4 + 16 q_0 + 25 q_0^2 + 14 q_0^3 - s_0\bigr) + 60 s_0 + 105 q_0 s_0 + 60 q_0^2 s_0 \Bigr) \Bigr]\,.
\end{eqnarray*}
Further we calculated the expression $H(y_1)$, as
\begin{eqnarray*}
    H(y_1) & = & H_0 \cdot \Bigl[ 1+ y_1 \cdot \Bigl( 1 + q_0 \Bigr) + \frac{y_1^2}{2} \cdot \Bigl( 2 + j_0 + 2 q_0 - q_0^2 \Bigr) +
    \frac{y_1^3}{6} \cdot \Bigl( -6 - 6 q_0 + 3 q_0^2 - 3 q_0^3 + j_0 \bigl(-3 + 4 q_0\bigr) + s_0 \Bigr) + \nonumber\\
    &&+\, \frac{y_1^4}{24} \cdot \Bigl( 24 - 4 j_0^2 + l_0 + 24 q_0 - 12 q_0^2 + 12 q_0^3 - 15 q_0^4 +
    j_0 \bigl(12 - 16 q_0 + 25 q_0^2\bigr) - 4 s_0 + 7 q_0 s_0 \Bigr) + \nonumber\\
    &&+\, \frac{y_1^5}{120} \cdot \Bigl( -120 + m_0 + j_0^2 \bigl(20 - 70 q_0\bigr) - 120 q_0 + 60 q_0^2 - 60 q_0^3 +
    75 q_0^4 - 105 q_0^5 + l_0 \bigl(-5 + 11 q_0\bigr) + \nonumber\\
    &&+\, 20 s_0 - 35 q_0 s_0 +
    60 q_0^2 s_0 + 5 j_0 \bigl(16 q_0 - 25 q_0^2 + 42 q_0^3 - 3 (4 + s_0)\bigr) \Bigr) \Bigr]\,,
\end{eqnarray*}
and in terms of the third redshift, $H(y_4)$,
\begin{eqnarray*}
     H(y_4) & = & H_0 \cdot \Bigl[ 1+ y_4 \cdot \Bigl( 1 + q_0 \Bigr) + \frac{y_4^2}{2} \cdot \Bigl( j_0 - q_0^2 \Bigr) +
    \frac{y_4^3}{6} \cdot \Bigl( -2 + 3 j_0 - 2 q_0 + 4 j_0 q_0 - 3 q_0^2 - 3 q_0^3 + s_0 \Bigr) + \nonumber\\
    &&+\, \frac{y_4^4}{48} \cdot \Bigl( 8 + 32 j_0 - 8 j_0^2 + 2 l_0 + 8 q_0 + 16 j_0 q_0 - 40 q_0^2 +
    50 j_0 q_0^2 - 48 q_0^3 - 30 q_0^4 - 8 \bigl(1 + q_0\bigr) + \nonumber\\
    &&+\, 8 j_0 \bigl(1 + 6 q_0\bigr) +
    16 s_0 + 14 q_0 s_0 \Bigr) + \nonumber\\
    &&+\, \frac{y_4^5}{240} \cdot \Bigl( -222 - 15 j_0 - 40 j_0^2 + 10 l_0 + 2 m_0 - 222 q_0 + 400 j_0 q_0 -
    140 j_0^2 q_0 + 22 l_0 q_0 + 15 q_0^2 + \nonumber\\
    &&+\, 250 j_0 q_0^2 - 300 q_0^3 +
    420 j_0 q_0^3 - 150 q_0^4 - 210 q_0^5 + 60 \bigl(1 + j_0 + q_0 - q_0^2\bigr) +
    100 s_0 - 30 j_0 s_0 + \nonumber\\
    &&+\, 70 q_0 s_0 + 120 q_0^2 s_0 -
    5 \bigl(-26 + 16 j_0^2 - 4 l_0 - 26 q_0 + 39 q_0^2 + 36 q_0^3 + 60 q_0^4 -
    j_0 (39 + 48 q_0 + 100 q_0^2) + \nonumber\\
    &&-\, 12 s_0 - 28 q_0 s_0\bigr) \Bigr) \Bigr]\,.
\end{eqnarray*}

\section{Pressure and derivatives as function of redshifts}
\label{app:pressure}
In this section we give the expressions for the pressure in terms of the
three redshifts, and also its derivatives with respect to these redshifts,
evaluated at present cosmic time $t_0$, which is equivalent to $z=y1=y_4=0$. \\
The result for the pressure in terms of $z$ reads
\begin{eqnarray*}
P(z=0) &=& \frac{1}{3} H_0^2 \Bigl(-1 + 2 q_0\Bigr) \,, \\
\frac{dP}{dz}\Big|_{z=0} &=& \frac{2}{3} H_0^2 \Bigl(-1 + j_0\Bigr) \,, \\
\frac{d^2P}{dz^2}\Big|_{z=0} &=& -\frac{2}{3} H_0^2 \Bigl(1 + j_0 + 2 q_0 + j_0 q_0 + s_0\Bigr)  \,, \\
\frac{d^3P}{dz^3}\Big|_{z=0} &=&  \frac{2}{3} H_0^2 \Bigl(-j_0^2 + l_0 + j_0 q_0
\bigl(4 + 3 q_0\bigr) + \bigl(4 + 3 q_0\bigr) s_0\Bigr) \,, \\
\frac{d^4P}{dz^4}\Big|_{z=0} &=& \frac{2}{3} H_0^2 \Bigl(9 j_0^2 - 9 l_0 - m_0 - 16 j_0 q_0 + 10 j_0^2 q_0 - 6 l_0 q_0 -
   27 j_0 q_0^2 - 15 j_0 q_0^3 - 16 s_0 + 5 j_0 s_0 - 27 q_0 s_0 - 15 q_0^2 s_0\Bigr)  \,,
\end{eqnarray*}
expressed in terms of $y_1$ we have
\begin{eqnarray*}
P(y_1=0) &=& \frac{1}{3} H_0^2 \Bigl(-1 + 2 q_0\Bigr) \,, \\
\frac{dP}{dy_1}\Big|_{y_1=0} &=& \frac{2}{3} H_0^2 \Bigl(-1 + j_0\Bigr) \,, \\
\frac{d^2P}{dy_1^2}\Big|_{y_1=0} &=& -\frac{2}{3} H_0^2 \Bigl(3 + j_0 \bigl(-1 + q_0\bigr) + 2 q_0 + s_0\Bigr)  \,, \\
\frac{d^3P}{dy_1^3}\Big|_{y_1=0} &=& -\frac{2}{3} H_0^2 \Bigl(12 + j_0^2 - l_0 + 12 q_0 + j_0 \Bigl(2 - 3 q_0\bigr) q_0 + 2 s_0 -
   3 q_0 s_0\Bigr) \,, \\
\frac{d^4P}{dy_1^4}\Big|_{y_1=0} &=& -\frac{2}{3} H_0^2 \Bigl(60 + m_0 + j_0^2 \bigl(3 - 10 q_0\bigr) + 72 q_0 + l_0 \bigl(-3 + 6 q_0\bigr) +
   j_0 \bigl(12 + 4 q_0 - 9 q_0^2 + 15 q_0^3 - 5 s_0\bigr) + \\
   &&+\, 4 s_0 - 9 q_0 s_0 + 15 q_0^2 s_0\Bigr)  \,,
\end{eqnarray*}
and finally with respect to $y_4$, the results for the pressure and its derivatives are
\begin{eqnarray*}
P(y_4=0) &=& \frac{1}{3} H_0^2 \Bigl(-1 + 2 q_0\Bigr) \,, \\
\frac{dP}{dy_4}\Big|_{y_4=0} &=& \frac{2}{3} H_0^2 \Bigl(-1 + j_0\Bigr) \,, \\
\frac{d^2P}{dy_4^2}\Big|_{y_4=0} &=& -\frac{2}{3} H_0^2 \Bigl(1 + j_0 + 2 q_0 + j_0 q_0 + s_0\Bigr) \,, \\
\frac{d^3P}{dy_4^3}\Big|_{y_4=0} &=& -\frac{2}{3} H_0^2 \Bigl(2 - 2 j_0 + j_0^2 - l_0 - 4 j_0 q_0 - 3 j_0 q_0^2 - 4 s_0 -
   3 q_0 s_0\Bigr) \,, \\
\frac{d^4P}{dy_4^4}\Big|_{y_4=0} &=& \frac{1}{3} H_0^2 \Bigl(-24 - 8 \bigl(-1 + j_0\bigr) - 11 j_0 + 18 j_0^2 - 18 l_0 - 2 m_0 -
   32 q_0 - 28 j_0 q_0 + 20 j_0^2 q_0 - 12 l_0 q_0 - 18 j_0 q_0^2 + \\
   &&-\, 30 j_0 q_0^3 + j_0 \bigl(3 - 20 q_0 - 36 q_0^2\bigr) - 48 s_0 + 10 j_0 s_0 -
   54 q_0 s_0 - 30 q_0^2 s_0\Bigr) \,.
\end{eqnarray*}

\section{Luminosity distance as function of the EoS parametrization}
\label{app:EoS}
These are the expressions for the luminosity distance $d_L$ in terms not of the CS,
but in terms of the pressure and its derivatives with respect to redshift.
To make the equations more readable, we introduce the notions $P_1 = \frac{dP}{dy_i}$,
$P_2 = \frac{d^2P}{dy_i^2}$ etc., where $y_i = z,y_1,y_4$. \\
For redshift $z$ we have
\begin{eqnarray*}
 d_L(z) & = & \frac{1}{H_0} \cdot \Bigl[  z + \frac{z^2}{4} \cdot \Bigl( 1-3\omega \Bigr)
   + z^3 \cdot \Bigl( -\frac{1}{8} - \frac{P_1}{4 H_0^2} + \omega + \frac{9 \omega^2}{8} \Bigr) + \nonumber\\
 &&+\, \frac{z^4}{64 H_0^2} \cdot \Bigl( -4 P_2 + P_1\cdot \left(34 + 54 \omega\right) - 5 H_0^2 \left(-1 +
 17 \omega + 45 \omega^2 + 27 \omega^3 \right) \Bigr) + \nonumber\\
 &&+\, \frac{z^5}{640 H_0^4} \cdot \Bigl( 108 P_1^2 + 5 H_0^4 (-7 + 218 \omega + 1008 \omega^2 +
  1350 \omega^3 + 567 \omega^4) + \nonumber\\
 &&-\, 4 H_0^2 \left(-26 P_2 + 2 P_3 - 36 P_2 \omega + P_1 \left(136 + 531 \omega
  + 405 \omega^2\right)\right) \Bigr) + \nonumber\\
 &&+\, \frac{z^6}{7680 H_0^4} \cdot \Bigl( -45 H_0^4 (1 + \omega)^2 \left(-7 + 375 \omega + 1827 \omega^2 +
  1701 \omega^3\right) + 4 H_0^2 \cdot 9 P_1 (257 + 1817 \omega + 3135 \omega^2 + \nonumber\\
  &&+\, 1575 \omega^3 ) - 36 P_1 \left(-20 P_2 + P_1 (169 + 225 \omega)\right) -
  8 H_0^2 \left[2 P_4 - 5 P_3 (7 + 9 \omega) + 9 P_2 \left(31 + 106 \omega + 75 \omega^2\right)\right] \Bigr) \Bigr]\,,
\end{eqnarray*}
for redshift $y_1$ the luminosity distance is given by
\begin{eqnarray*}
 d_L(y_1) & = & \frac{1}{H_0} \cdot \Bigl[  y_1 + \frac{y_1^2}{4} \cdot \Bigl( 5-3\omega \Bigr)
      + \frac{y_1^3}{8} \cdot \Bigl(11 - \frac{2 P_1}{H_0^2} - 4 \omega + 9 \omega^2 \Bigr) + \nonumber\\
 &&-\, \frac{y_1^4}{64 H_0^2} \cdot \Bigl( 4 P_2 + P_1 (6 - 54 \omega) + H_0^2 (-93 + 37 \omega + 9 \omega^2 +
    135 \omega^3) \Bigr) + \nonumber\\
 &&+\, \frac{y_1^5}{640 H_0^4} \cdot \Bigl( 108 P_1^2 + 5 H_0^4 (193 - 78 \omega + 72 \omega^2 + 270 \omega^3 + 567 \omega^4) -
      4 H_0^2 (2 (P_2 + P_3 - 18 P_2 \omega) + \nonumber\\
 &&+\, P_1 (20 + 63 \omega + 405 \omega^2)) \Bigr) + \nonumber\\
 &&+\, \frac{y_1^6}{7680 H_0^4} \cdot \Bigl( -15 H_0^4 (-793 + 323 \omega - 210 \omega^2 + 990 \omega^3 +
       4347 \omega^4 + 5103 \omega^5) + 36 P_1 (-20 P_2 + P_1 (29 + \nonumber\\
 &&+\, 225 \omega)) + 4 H_0^2 (3 P_1 (-85 + 243 \omega + 2205 \omega^2 + 4725 \omega^3) - 2 (P_3 + 2 P_4 - 45 P_3 \omega +
      3 P_2 (7 + 48 \omega + 225 \omega^2))) \Bigr) \Bigr]\,,
\end{eqnarray*}
and for the last redshift $y_4$ the result reads
\begin{eqnarray*}
 d_L(y_4) & = & \frac{1}{H_0} \cdot \Bigl[  y_4 + \frac{y_4^2}{4} \cdot \Bigl( 1-3\omega \Bigr)
    + y_4^3 \cdot \Bigl(  \frac{5}{24} - \frac{P_1}{4 H_0^2} + \omega + \frac{9 \omega^2}{8}\Bigr) \nonumber\\
 &&+\, \frac{y_4^4}{192 H_0^2} \cdot \Bigl( -H_0^2 (-47 + 351 \omega + 675 \omega^2 + 405 \omega^3) +
    6 \left(-2 P_2 + P_1 (17 + 27 \omega)\right) \Bigr) \nonumber\\
 &&+\, \frac{y_4^5}{1920 H_0^4} \cdot \Bigl( 324 P_1^2 + H_0^4 (-89 + 5190 \omega +
    17280 \omega^2 + 20250 \omega^3 + 8505 \omega^4) + \nonumber\\
 &&-\, 12 H_0^2 \left( -26 P_2 + 2 P_3 - 36 P_2 \omega + P_1 (172 + 531 \omega + 405 \omega^2) \right) \Bigr) + \nonumber\\
 &&+\, \frac{y_4^6}{23040 H_0^4} \cdot \Bigl( H_0^4 (-5521 + 96063 \omega + 454950 \omega^2 + 838350 \omega^3 +
    705915 \omega^4 + 229635 \omega^5) + \nonumber\\
 &&-\, 12 H_0^2 \left[P_1 (3533 + 18333 \omega + 28215 \omega^2 + 14175 \omega^3) - 2 \left(2 P_4 -
    5 P_3 (7 + 9 \omega) + P_2 (343 + 954 \omega + 675 \omega^2)\right)\right] + \nonumber\\
 &&+\, 108 P_1 \left(-20 P_2 + P_1 (169 + 225 \omega)\right) \Bigr) \Bigr]\,.
\end{eqnarray*}

\section{Derivatives of G(z)}
\label{app:G}
Here we report the results for the derivatives of the function $G(z)$, which characterizes
the equation of state of dark energy in a specific model, evaluated at present time, for
the three redshifts under consideration. \\
The derivatives with respect to redshift $z$ at $z=0$ read
\begin{eqnarray*}
\frac{dG}{dz}\Big|_{z=0} &=& 2 - 3 \Omega_m + 2 q_0 \,, \\
\frac{d^2G}{dz^2}\Big|_{z=0} &=& -2 \left(-1 - j_0 + 3 \Omega_m - 2 q_0\right)  \,, \\
\frac{d^3G}{dz^3}\Big|_{z=0} &=& -2 \left(3 \Omega_m + j_0 q_0 + s_0\right)  \,, \\
\frac{d^4G}{dz^4}\Big|_{z=0} &=& 2 \bigl(-12 j_0 + 3 j_0^2 + 12 j_0 \Omega_m - 4 j_0^2 \Omega_m + l_0 \Omega_m - 28 j_0 q_0 +
   32 j_0 \Omega_m q_0 + 12 q_0^2 - 22 j_0 q_0^2 + \\
  &&-\, 12 \Omega_m q_0^2 + 25 j_0 \Omega_m q_0^2 + 24 q_0^3 - 24 \Omega_m q_0^3 + 15 q_0^4 -
  15 \Omega_m q_0^4 - 4 s_0 + 8 \Omega_m s_0 - 4 q_0 s_0 + \\
  &&+\, 7 \Omega_m q_0 s_0 + (1 - \Omega_m) (-4 j_0^2 + l_0 - 12 q_0^2 - 24 q_0^3 - 15 q_0^4 +
       j_0 (12 + 32 q_0 + 25 q_0^2) + 8 s_0 + 7 q_0 s_0)\bigr)  \,,\\
\frac{d^5G}{dz^5}\Big|_{z=0} &=& -2 \bigl(10 l_0 + m_0 + 6 l_0 q_0 - 10 j_0^2 (1 + q_0) +
   5 j_0 (4 q_0 + 6 q_0^2 + 3 q_0^3 - s_0) + 20 s_0 + 30 q_0 s_0 + 15 q_0^2 s_0 \bigr)  \,,
\end{eqnarray*}
expressed in terms of $y_1$ we have at $y_1=0$
\begin{eqnarray*}
\frac{dG}{dy_1}\Big|_{y_1=0} &=& 2 - 3 \Omega_m + 2 q_0 \,, \\
\frac{d^2G}{dy_1^2}\Big|_{y_1=0} &=& -2 (-2 - j_0 + 6 \Omega_m - 2 q_0 + q_0^2 - (1 + q_0)^2)  \,, \\
\frac{d^3G}{dy_1^3}\Big|_{y_1=0} &=& -2 (-6 + 30 \Omega_m - 6 q_0 + 3 q_0^2 - 3 q_0^3 + j_0 (-3 + 4 q_0) -
   3 (1 + q_0) (2 + j_0 + 2 q_0 - q_0^2) + s_0) \,,
\end{eqnarray*}
and finally with respect to $y_4$, the results for the derivatives at present time $y_4=0$ are
\begin{eqnarray*}
\frac{dG}{dy_4}\Big|_{y_4=0} &=& 2 - 3 \Omega_m + 2 q_0 \,, \\
\frac{d^2G}{dy_4^2}\Big|_{y_4=0} &=& 2 \left(1 + j_0 - 3 \Omega_m + 2 q_0\right)  \,, \\
\frac{d^3G}{dy_4^3}\Big|_{y_4=0} &=& -2 \left(-2 + 6 \Omega_m + (-2 + j_0) q_0 + s_0\right)  \,, \\
\frac{d^4G}{dy_4^4}\Big|_{y_4=0} &=& 2 \left(8 - j_0^2 + l_0 - 24 \Omega_m + 16 q_0 +
    j_0 (8 + 4 q_0 + 3 q_0^2) + 4 s_0 + 3 q_0 s_0\right)  \,,\\
\frac{d^5G}{dy_4^5}\Big|_{y_4=0} &=& -2 \bigl(-16 + m_0 + 84 \Omega_m - 16 q_0 - 10 j_0^2 (1 + q_0)
    + 2 l_0 (5 + 3 q_0) + 5 j_0 (8 q_0 + 6 q_0^2 + 3 q_0^3 - s_0) + \\
  &&+\, 40 s_0 + 30 q_0 s_0 + 15 q_0^2 s_0 \bigr) \,.
\end{eqnarray*}

\end{widetext}
\end{document}